\documentclass[aip, jcp, amsmath,amssymb, reprint]{revtex4-1}

\usepackage{amsmath}
\usepackage{amssymb}
\usepackage{graphicx}
\usepackage{siunitx}

\usepackage[table]{xcolor}
\usepackage{array,tabularx}

\usepackage{gensymb}

\newenvironment{conditions*}
{\par\vspace{\abovedisplayskip}\noindent
	\tabularx{\columnwidth}{>{}l<{} @{}>{${}}c<{{}$}@{} >{\raggedright\arraybackslash}X}}
{\endtabularx\par\vspace{\belowdisplayskip}}

\usepackage{listings}

\usepackage{booktabs}

\begin{document}
	
	
\title{Extended Wertheim theory predicts the anomalous chain length distributions of divalent patchy particles under extreme confinement}

	\author{H.J. Jonas}
	\affiliation{van 't Hoff Institute for Molecular Sciences, University of Amsterdam,  PO Box 94157,
		1090 GD Amsterdam, The Netherlands}%
		\author{P. Schall}
	\affiliation{Van der Waals-Zeeman Institute, Institute of Physics, University of Amsterdam, PO Box 94485,   1090 GL Amsterdam, The Netherlands}%
	\author{P.G. Bolhuis}%
	\email{p.g.bolhuis@uva.nl}
	\affiliation{van 't Hoff Institute for Molecular Sciences, University of Amsterdam,  PO Box 94157, 1090 GD Amsterdam, The Netherlands}%
	
	\date{\today}

	\begin{abstract}
		Colloidal patchy particles with divalent attractive interaction can   self-assemble into linear polymer  chains. Their equilibrium properties in 2D and  3D are well described
		by 
		Wertheim's  thermodynamic perturbation theory which predicts a well-defined exponentially decaying equilibrium chain length distribution. 
		In experimental realizations, due to gravity, particles  sediment to the bottom of the suspension forming a monolayer of particles with a gravitational height smaller than  the particle diameter. 
		 In accordance with experiments, an  anomalously high monomer concentration is observed in simulations which is not well understood. 
		To  account for this observation,  we interpret the polymerization as taking place in a highly confined  quasi-2D plane and  extend the Wertheim thermodynamic perturbation theory  by defining addition reactions constants as functions of the chain length. 
		We derive the theory, test it on simple square well potentials, 
		and apply it to the experimental case  
		of synthetic colloidal patchy particles immersed in a binary liquid mixture that are described by  an accurate effective  critical Casimir patchy particle potential. The important  interaction parameters  entering the theory are  explicitly computed using   the integral method in combination with Monte Carlo sampling.
			Without any adjustable parameter, the predictions of the chain length distribution are in excellent agreement with explicit simulations of self-assembling particles. 
			We discuss generality of the approach, and its application range. 
	\end{abstract}
	\maketitle

	Synthetic  colloidal particles suspended in a  near-critical binary liquid mixture (e.g. water and lutidine),  attract each other  via a solvent mediated critical Casimir force.   Through novel synthesis routes these particles can be designed such that they form  directed bonds between 
	patches on  the surface of neighboring particles \cite{Gong2017}.
	As such patchy particles simultaneously  experience thermal motion,  their statistical behavior follows the  Boltzmann distribution. Hence, they can be viewed as  mesoscopic analogs of (carbon) atoms, which can be  directly observed via, e.g., confocal microscopy\cite{Swinkels2021}. In this way, patchy particles can  act as  an experimental model system to explore complex self-assembled structures analogous to molecular architectures, such as chains, rings, and networks \cite{Wang2012,Nguyen2017a,Stuij2021a,Swinkels2021,Stuij2021}.
        
To understand the self-assembly in patchy particle systems, one can of course resort to computer simulations\cite{Wolters2015,Avvisati2015,Roldan-Vargas2017,Newton2017a,Newton2015}, but an attractive alternative is to invoke  statistical mechanics which aids to a better theoretical understanding and prediction. One of the classical theories for self-assembly of colloidal particles is the Wertheim thermodynamic perturbation theory (TPT)\cite{Wertheim1984,Wertheim1984a,Wertheim1986,Wertheim1986a}, later reformulated as Statistical Associating Fluid Theory  (SAFT) by Chapman et al.~\cite{Chapman1989}.
	Wertheim's theory was originally intended as a molecular model \cite{Chapman1990}, 
	 but also works for mesoscopic particles. For divalent patchy particles in two and three dimensions, 
	Wertheim theory is able to predict the polymerization equilibrium  in terms of for example the chain length distribution, with  the total particle density  and a  pair bonding strength  as the only input parameters \cite{Sciortino2007,Tartaglia2010,Russo2010,Rovigatti2013,Phys2016,Stopper2020}.  
	 For systems with average valencies larger than two, equilibrium properties  are predicted using Flory-Stockmayer's polymer theory\cite{Stockmayer1943,Sciortino2011}.  The location of the percolation point, existence of  empty liquids and equilibrium gels were predicted theoretically, confirmed in simulation and validated experimentally\cite{Oleksy2015,Bianchi2007a,Teixeira2017,Sciortino2017,Biffi2013,Ruzicka2011}.

However, when there is a mismatch in density of the particles and the suspending solvent,  particles will sediment to the bottom of the sample due to gravity. 
For sufficiently low particle concentration (or volume fraction) and short gravitational height, the system is then confined to a quasi-2D plane; making  single layer structures possible. 
Direct application of Wertheim's theory for divalent particles  in 2D or 3D will  give an exponential distribution and  a large discrepancy between the experiments and theoretical prediction exists, in particular in the monomer (and dimer) density\cite{Jonas2021,Stuij2021}. In this work we address this discrepancy.

\begin{figure*}[t!]
	\centering
	\includegraphics[width=14cm]{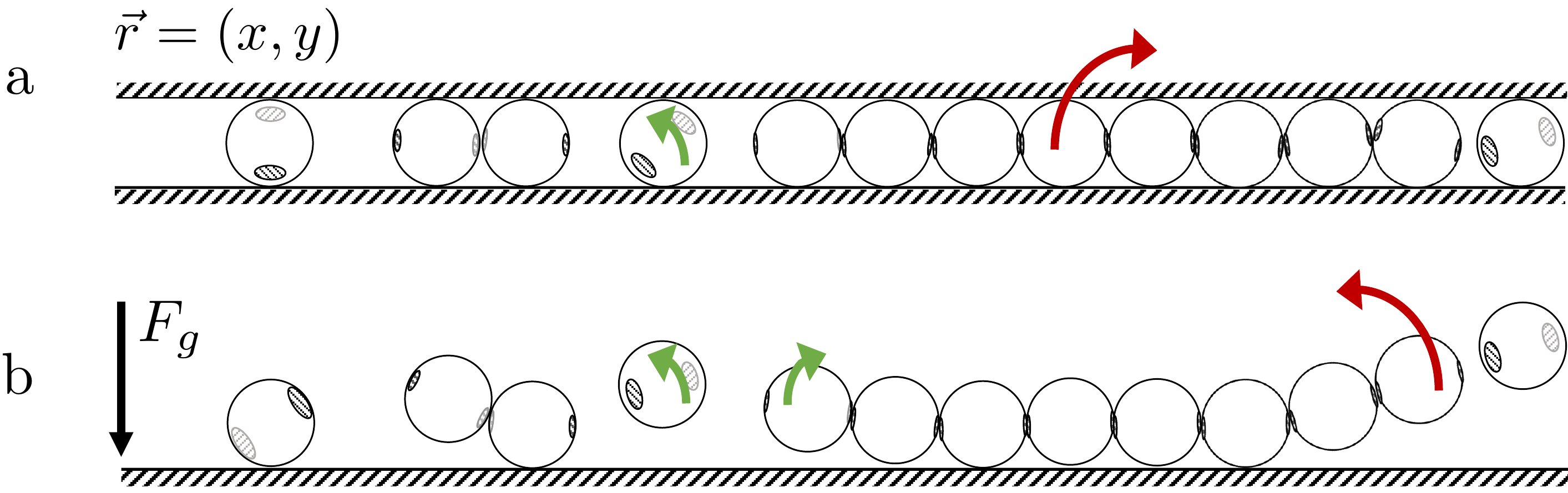}
	\caption{Schematic illustration of possible patchy particle orientations in quasi-2D or under the gravitational field confining the particles close to the wall (striped area). (a) In quasi-2D, i.e. with the translation restricted to the $(x,y)-$plane,  chains cannot rotate around their center-of-mass against the confining wall (red arrow) while monomers can (green arrow). (b) At finite gravitational field, short chains have more freedom to translate against the gravitational force $F_g$ compared to long chains due to the stiffness of the bonds.
	} 
	\label{pic_sideview_box}
\end{figure*}

The  origin of the discrepancy is that, under extreme confinement where spherical particles live in a two-dimensional $(x,y)-$plane,  the monomers are still able to rotate around their center-of-mass (Fig.~\ref{pic_sideview_box}a).  While some monomer orientations, e.g. when their patch points toward the wall,  are part of the orientational phase space, their patches are not available for bonding. This renders the system fundamentally  different compared to the standard 2D and 3D systems. As a consequence, in order to predict thermodynamic properties,  this excess rotational degree of freedom needs to be taking into account \cite{Sciortino2019a}.
For strong confinement due to gravity, there is an additional anisotropy in the density along the direction perpendicular to the wall.  Monomers and small chains still having freedom to translate and rotate against gravity, while for long chains only a small part of the chain (at the end) has this freedom as illustrated in Fig.~\ref{pic_sideview_box}b. 

Note that directional assembly under extreme confinement not only occurs in model colloidal patchy particle systems.
In chemistry, there are many examples where confinement has been used at its advantage. For example, nanoporous materials with pore shapes and sizes comparable to the typical size of small molecules such as metal-organic-frameworks (MOFs), covalent organic frameworks (COFs), zeolitic imidazolate frameworks (ZIFs)\cite{Li2017,Dubbeldam2019}, and nanopores composed of for example carbon nanotubes\cite{Radha2016}. 
Another example is that of  self-assembled supramolecular structures where the intermolecular non-covalent bonds determine the structure and chemical function, and confinement affects the reactivity\cite{Mouarrawis2018}. Even in the confined environment of the living cell, where short-ranged, strongly directional hydrogen bonds provide a mode for molecular  assembly.
  Recently, the fabrication of nanoslits with \AA ngstr\"{o}m-scale separation became possible  opening an exciting field of nanofluidics that show unusual dynamics, kinetics, and thermodynamics due to the extreme confinement\cite{Esfandiar2017,Munoz-Santiburcio2017,Bocquet2020,Seo2021}. However, a thorough theoretical understanding of the effect of confinement on for example  separation and phase transitions is still lacking\cite{Thompson2018,Faucher2019}. 

Such  highly confined systems could be in principle described theoretically with TPT\cite{Sokoowski2014,Marshall2016,BrazTeixeira2021}. When solving TPT, one has to compute the interaction parameters in the theory. There are two major routes to do this: 
via an 'integral method'\cite{Chapman1988} or via classical density functional theory (DFT) \cite{Kierlik1992,Kierlik1993,Kierlik1994}.
Solving TPT  becomes increasingly complex for inhomogeneous systems due to positional and orientational coupling\cite{Teixeira2019}. While extremely powerful, the DFT-route is currently  not able to predict thermodynamic equilibrium for the  Wertheim theory  in highly confined systems at low temperatures quantitatively. See e.g. Ref.~\onlinecite{CamachoVergara2020}, which shows excellent predicted density distributions for tetrapatch particles at large wall separation, but for small wall separations (between $1.18-3.02$ times the particle diameter) shows discrepancies for three different associating density functionals. In contrast, Ref.~\onlinecite{Marshall2015}, following the integral route, predicted densities accurately of spherical dipatch particles in a one-dimensional pore with a width of the particle diameter. 

	In this paper we take a different approach. 
	We interpret the rotational and translational freedom against the gravitational field as an additional source of entropy. This orientation-positions-dependent entropy effectively reduces the reactivity.
	We separate the polymerization reactions of the species which
	  gives rise to adapted expressions for the chain length distributions in the Wertheim theory. 
 By computing  the interaction parameter via the integral method  with Monte Carlo (MC) sampling, we can directly calculate the excess rotational and translational entropy and capture the corresponding equilibrium reaction constants between the species. 
	
To validate our approach, we simulate patchy particles interacting via a simple square well potential combined with different forms of orientation-dependent switching functions under various gravitational strengths. 
Additionally, we apply the theory to the critical Casimir dipatch colloid  particle system, for which the chain length distribution was experimentally studied in Ref.~\onlinecite{Stuij2021} and an accurate effective potential model was developed recently in Ref.~\onlinecite{Jonas2021}.
The extended Wertheim theory contains no fit parameters, and needs only two input parameters: the species-dependent interaction parameters and the total particle number density $\rho$.
We compare the predicted distribution with 
 the simulated ones, and find excellent quantitative agreement.

The paper is organized as follows:  we start with a brief overview of the traditional Wertheim theory that holds both for 2D and 3D.  
Then, we will introduce the adapted Wertheim theory for the highly confined system in quasi-2D 
followed by the gravitationally confined systems.
Using the quasi-2D systems, we show how we can determine the excess rotational free energy of the monomers and its effect on the chain length distribution. 
Next, we introduce the external gravitational field which gives also short chains additional entropy and thus higher probability of occurrence and show that the flexibility of the chain also plays a role on the distributions. This effect too can be determined via the integral method giving excellent predictions of the chain length distributions of divalent colloidal particles. Finally, we apply and validate the theory on our accurate patchy particle model interacting via critical Casimir interactions under realistic gravitational conditions.
	We end with concluding remarks, and  a future outlook.

	\section{Theory}
	\subsection{First order thermodynamic perturbation theory \label{sec:TPT1}}
       
	
	
	Consider a 3D suspension of  hard spherical particles or a 2D suspension of hard disks, in which each particle is divalent, i.e dressed with two attractive patches, usually located at opposing poles.  Each patch or site is able to make a bond with a site on another particle, resulting in the  association of particles into linear chains.  Moreover, each site is able to make only one single bond, and each bond is equally likely to form.   The aggregation of the monomers into larger clusters can  then be viewed as a set of addition reactions: 
	\begin{subequations}
	\begin{align}
		A_1 + A_1 \rightleftharpoons A_2 \\
		A_1 + A_2 \rightleftharpoons A_3 \\
		\dots \notag \\
		A_1 + A_{n-1} \rightleftharpoons A_n 
	\end{align}
	\end{subequations}
	where $A_1$ stands for a monomer, $A_2$ for a dimer, and $A_n$ for a chain composed of $n$ monomers. This type of reactivity is also known as isodesmic polymerization where each addition reaction of a monomer is associated with equal amount of free energy\cite{SCHOOT200945}. All reactions have an equilibrium constant $K$, defined through the law of mass action:
	\begin{align}
		K  & = \frac{[A_n]}{[A_1] [A_{n-1}]}  \notag \\
		& = \frac{\rho_n}{\rho_1 \rho_{n-1}},  \label{K} 
	\end{align}
	where $[A_n]$ denotes the concentration of $A_n$, and $\rho_n$ the number density of $n$-mers, i.e., chains of length $n$.
	
	For this situation one can
	apply Wertheim 's first-order thermodynamic perturbation theory (TPT1),  and calculate the chain length probability distribution\cite{Wertheim1984,Sciortino2007}. The probability of observing a non-occupied binding site is denoted as $X$. Then the number fraction of chains of size $n$ is on average given by
	\begin{equation}
		\rho_n/\rho  = X^2 (1-X)^{n-1}, \label{rho_n_rho}
	\end{equation}
	with $\rho$ the number density. This expression is rationalised as follows. A cluster of size $n$ has $(n-1)$ links. The probability of forming a link is $1-X$. The probability of forming $n-1$ links is thus $(1-X)^{n-1}$.  However, as there are two unoccupied reactive sites, that accounts for a factor $X^2$. So, for a monomer this reduces to $\rho_1= \rho X^2$. To convince oneself that this is consistent, one can add up all chain lengths, which would have to add up to the total density of particles:
	\begin{equation}
\label{eq:WTtotdens}
          \rho = \sum_n n\rho_n =  \rho X^2 \sum_n n(1-X)^{n-1}.
	\end{equation}
	The geometric sum adds up to $1/X^2$, which is indeed consistent. 
	
	
	Next, using Eq.~\ref{rho_n_rho} and $\rho_1= \rho X^2$,  the equilibrium  reaction constant $K$ from Eq.~\ref{K} is rewritten as:
	\begin{align}
		\label{eq:WTK}
		K 
		=&\frac{1-X}{\rho X^2}. 
	\end{align}
	Solving for X gives 
	\begin{align}
		X = \frac{2}{1+\sqrt{1+4K\rho}}.  \label{X_wertheim}
	\end{align}
	Thus, given a density $\rho$ as well as an equilibrium constant $K$, Eq. ~\ref{rho_n_rho}  together with Eq. ~\ref{X_wertheim}  form a complete description of the system.

	The slope of (the log of) the chain length distribution is: 
	\begin{align}
		K \rho_1 & =  \frac{\rho_n}{\rho_{n-1}}  \notag \\
		&= 1-X 
		\equiv X_b, \label{eq:slopeprob}
	\end{align}
	where the latter equality defines the  probability $X_b$ for binding or, equivalently, the fraction of bound sites $X_b$.
	
	Traditionally, TPT computes the equilibrium constant $K$ via  the interaction parameter $\Delta$\cite{Chapman1988}, as $K \equiv \left<M\right>\Delta$, with $M$ the number of binding sites or patches  per  particle.
	The $\Delta$-parameter represents the (exponential of the) free energy difference of the bonding reaction with respect to the hard particle reference state. It is calculated via an integration over space of the Boltzmann weighted energy averaged over the allowed orientations of the particles, multiplied by the probability of finding a particle at distance $r$, i.e. the radial distribution function  $g(\mathbf{r})$. 
	The interaction parameter $\Delta$ is then:
	\begin{align}
		\Delta & =    \int g(\mathbf{r})  \left\langle f(r,\Omega_{\alpha},\Omega_{\gamma})\right\rangle _{\Omega_{\alpha},\Omega_{\gamma}} d\mathbf{r} 
		\label{Delta},
	\end{align}
	where $\mathbf{r}$ is the inter-particle vector of particle $\alpha$ and $\gamma$ 
	with their orientations $\Omega_{\alpha}$ and  $\Omega_{\gamma}$, respectively, $g(\mathbf{r})$ is the pair correlation function of the reference systems e.g. hard spheres, $f(r,\Omega_{\alpha},\Omega_{\gamma}) = \exp^{-\beta U(\mathbf{r}, \Omega_{\alpha},\Omega_{\gamma})} -1$ is the Mayer function, and, finally, $\left\langle \right\rangle _{\Omega_{\alpha},\Omega_{\gamma}}$ denotes the orientational average of the Mayer function  of particle $\alpha$ and $\gamma$ separated at distance $|\mathbf{r}|=r$.
	This calculation of $\Delta$ is non-trivial, and depends on the geometry of the setup. 	
	
	\subsection{Wertheim theory in quasi-2D  \label{q2D_TPT1}}

	 The treatment in Sec~\ref{sec:TPT1} assumes that all association reactions follow identical statistics. However,  the situation is slightly different when confining the chain formation to a plane by e.g. two walls with particle diameter separation (Fig.~\ref{pic_sideview_box}A). In principle, the above described TPT/SAFT framework also applies in that case, except for one crucial difference in the assumption about the reactivity. Upon binding of single particles (free monomers) an excess rotational entropy is lost. 
	Therefore, the first reaction in the series  where two free monomers react to form a dimer  (Eq.~\ref{2M_reaction})  is fundamentally different from the others where only one free monomer reacts with an existing cluster of size $n>1$  (Eq.~\ref{1M_reaction}-\ref{1M_reaction2}):
\begin{subequations}
	\begin{align}
		A_1 + A_1 \stackrel{\mathrm{K_{2}}}{\rightleftharpoons}  A_2  \label{2M_reaction}\\
		A_1 + A_2 \stackrel{\mathrm{K}}{\rightleftharpoons} A_3 \label{1M_reaction} \\
		\dots  \notag \\
		A_1 + A_{n-1} \stackrel{\mathrm{K}}{\rightleftharpoons}  A_n \label{1M_reaction2} 
	\end{align}
	\end{subequations}
	 which leads to an increased monomer concentration as observed in the chain length distribution. 	
	 
\begin{figure}[t!]
		\centering		
		\includegraphics[height=2.7cm]{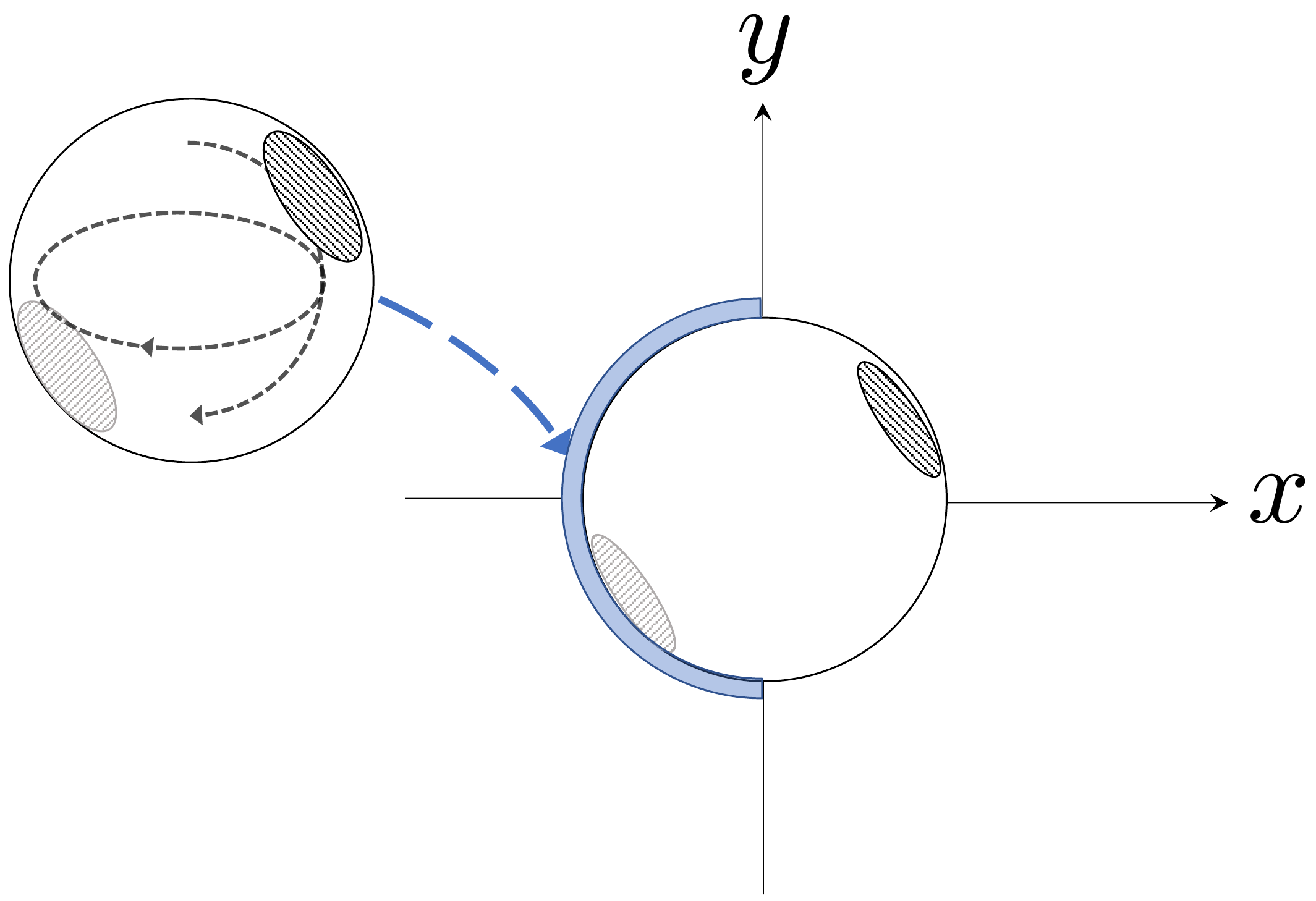}	\includegraphics[height=2.7cm]{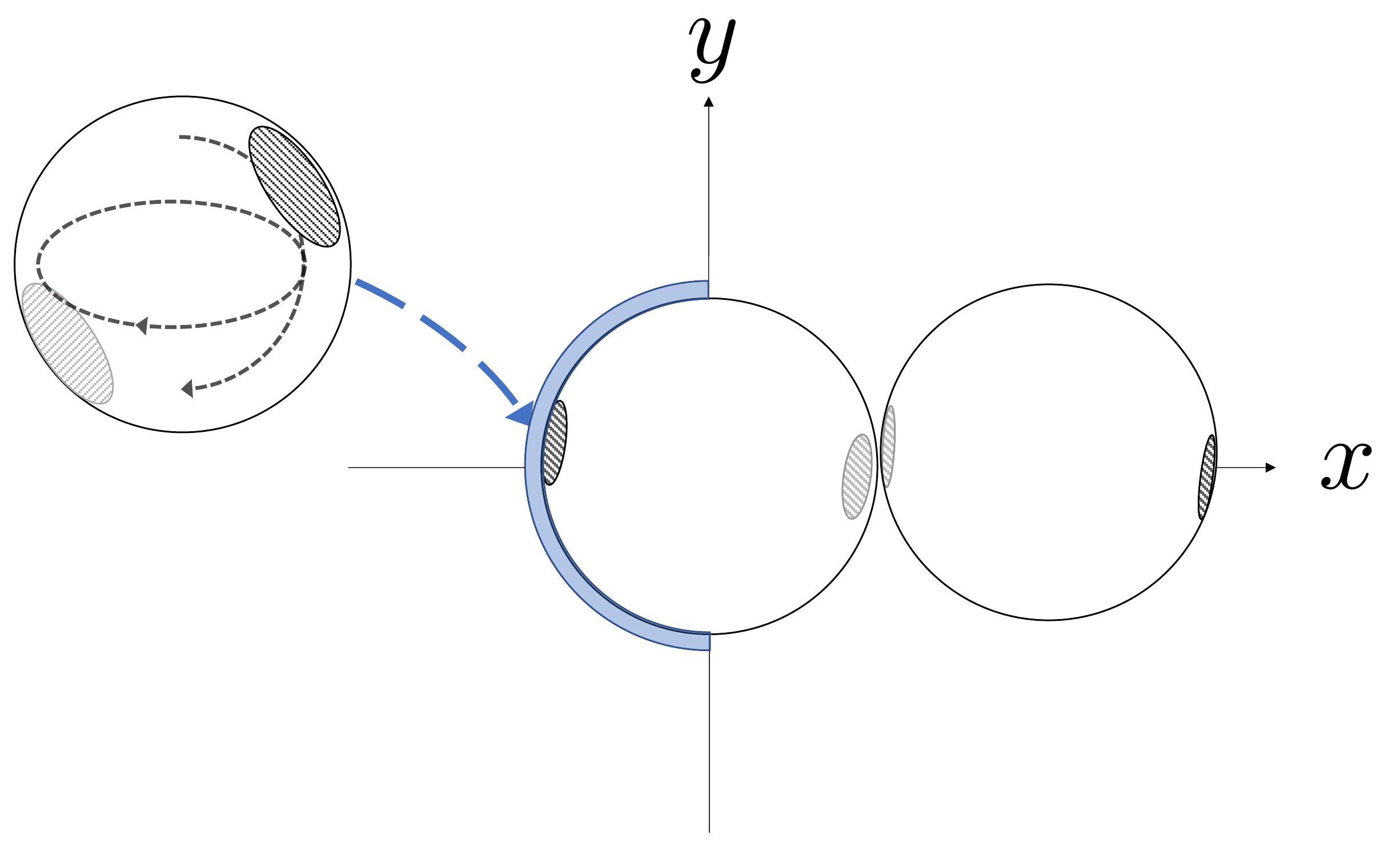}
		\caption{Schematic illustration of the two differences in sampling $K_2$ and $K$ in the left and right picture, respectively. The freely rotating $\alpha$-particle (with dotted curved arrows indicate rotation axes) is radially sampled uniformly in the integration area $V$ (blue area) around the $\gamma$-particle which is positioned at $(0,0)$. Due to symmetry we can reduce the sampling volume to a hemispherical shell around the $\gamma$-particle. } 
		\label{q2D_MC_sampling}
	\end{figure}

	By classifying not just one type of bonding reaction, but two types of reactions (see Fig. \ref{q2D_MC_sampling}), with corresponding free energies and equilibrium constants, we derive  the extended Wertheim theory in quasi-2D. For the two addition reactions, the constants $\rm K_2$ and $\rm K$  are given by 

	\begin{align}
		K_2 &= \frac{\rho_2}{\rho_1^2}, \notag \\
		K &= \frac{\rho_n}{\rho_1 \rho_{n-1}}.
	\end{align}
	Rewriting the first equation gives
	\begin{align}
		\rho_2=    K_2  \rho_1^2 =   \rho_1 (K_2  \rho_1)
	\end{align}
	while the next addition reaction yields
	\begin{align}
		\rho_3=    K \rho_1 \rho_2    = \rho_1 (K_2  \rho_1)   (K \rho_1).
	\end{align}
Continuing along this line, for $n\ge 2$ it follows that
\begin{align}
  \label{eq:WTq2D_rhon}
		\rho_n=    \rho_1 (K_2  \rho_1)   (K \rho_1)^{n-2}.
	\end{align}
	Note that the above equations express  the number densities of the clusters. 
	So adding all densities multiplied by the chain lengths will give the imposed total particle density $\rho$
	\begin{align}
		\rho = \rho_1 + \sum_{n=2}n \rho_n =     \rho_1 + \sum_{n=2} n \rho_1 (K_2  \rho_1)   (K \rho_1)^{n-2}.
	\end{align}
	Taking out a factor $K_2/K$ from the sum leads to
	\begin{align}
		\rho =   \rho_1 + \frac{K_2}{K} \rho_1  \sum_{n=2} n (K \rho_1)^{n-1}.
	\end{align}
	To make this a tractable sum, we add and subtract a term  $\frac{K_2}{K} \rho_1$, yielding
	\begin{align}
          \label{eq:WTq2Dsum}
		\rho =   \rho_1 \left( 1-\frac{K_2}{K}\right)  + \frac{K_2}{K} \rho_1  \sum_{n=1} n  (K \rho_1)^{n-1}.
	\end{align}
When $K_2=K$, the first term vanishes on the rhs of Eq.~\ref{eq:WTq2Dsum}, recovering  Eq.~\ref{eq:WTtotdens} in the original TPT1. 
	In this we recover the fraction/probability of bound sites $X_b =1 -X = K \rho_1$ (see Eq.~\ref{eq:slopeprob}). 

	The geometric sum can now be evaluated, giving
	\begin{align}
             \label{eq:WTq2Dsum2}
		\rho =   \rho_1 \left( 1-\frac{K_2}{K}\right)  + \frac{K_2}{K} \rho_1 \frac{1}{(1- K \rho_1)^2}.
	\end{align}
	Next we divide by $\rho$, and define the monomer fraction $X_1 \equiv \rho_1/\rho$, yielding
	\begin{align}
		1 =   X_1 \left( 1-\frac{K_2}{K}\right) + \frac{K_2}{K} X_1 \frac{1}{(1- K \rho X_1)^2}. \label{X1_KK2}
	\end{align}
	
 Applying TPT1 amounts to solving Eq.~\ref{X1_KK2} for the unknown monomer fraction $X_1$, which in turn sets the entire distribution $\rho_n$  in Eq.~\ref{eq:WTq2Dsum}.  This  requires knowledge of the equilibrium constants $K$ and $K_2$ that are directly proportional to $\Delta$ in Eq.~\ref{Delta}, and follow from evaluating the integral. The difference between the calculations of the $\Delta$'s corresponding to $K$ and $K_2$ will be explained later in Sec.~\ref{sec:Delta_calculation}. 
 
 For the chain length probabilities $P_n$, the chain densities are normalized with the chain number density $\rho_c\equiv \sum_n \rho_n$, which, using Eq.~\ref{eq:WTq2D_rhon}, results, analogous to  Eq.~\ref{eq:WTq2Dsum2}, in
 \begin{equation}
 	\rho_c = \rho_1 \left(1-\frac{K_2}{K} \right) + \rho_1\frac{K_2}{K}\frac{1}{X}
 \end{equation}
 so that:
 \begin{align}
 	P_1 &= \rho_1/\rho_c  \notag \\
 	P_2  &= \rho_2/\rho_c \notag \\
 	&\dots \notag\\
 	P_n &= \rho_n/\rho_c  
 \end{align}
 In this way, also the average chain length follows:  $L=\sum_{n=1} nP_n = {\rho}/{\rho_c}$.

	\subsection{Wertheim in a gravitational field}
	
	The above description for quasi-2D confinement also holds for an infinitely short gravitational height, in which translation away from the confining wall is strongly suppressed.
	If the gravitational field is not so strong, the particles are able to levitate to the order of the gravitational height. In turn, this translational freedom affects both the reactivity of particle association, and the free energy. 

\begin{figure}[t!]
	\centering
	\includegraphics[height=2.65cm]{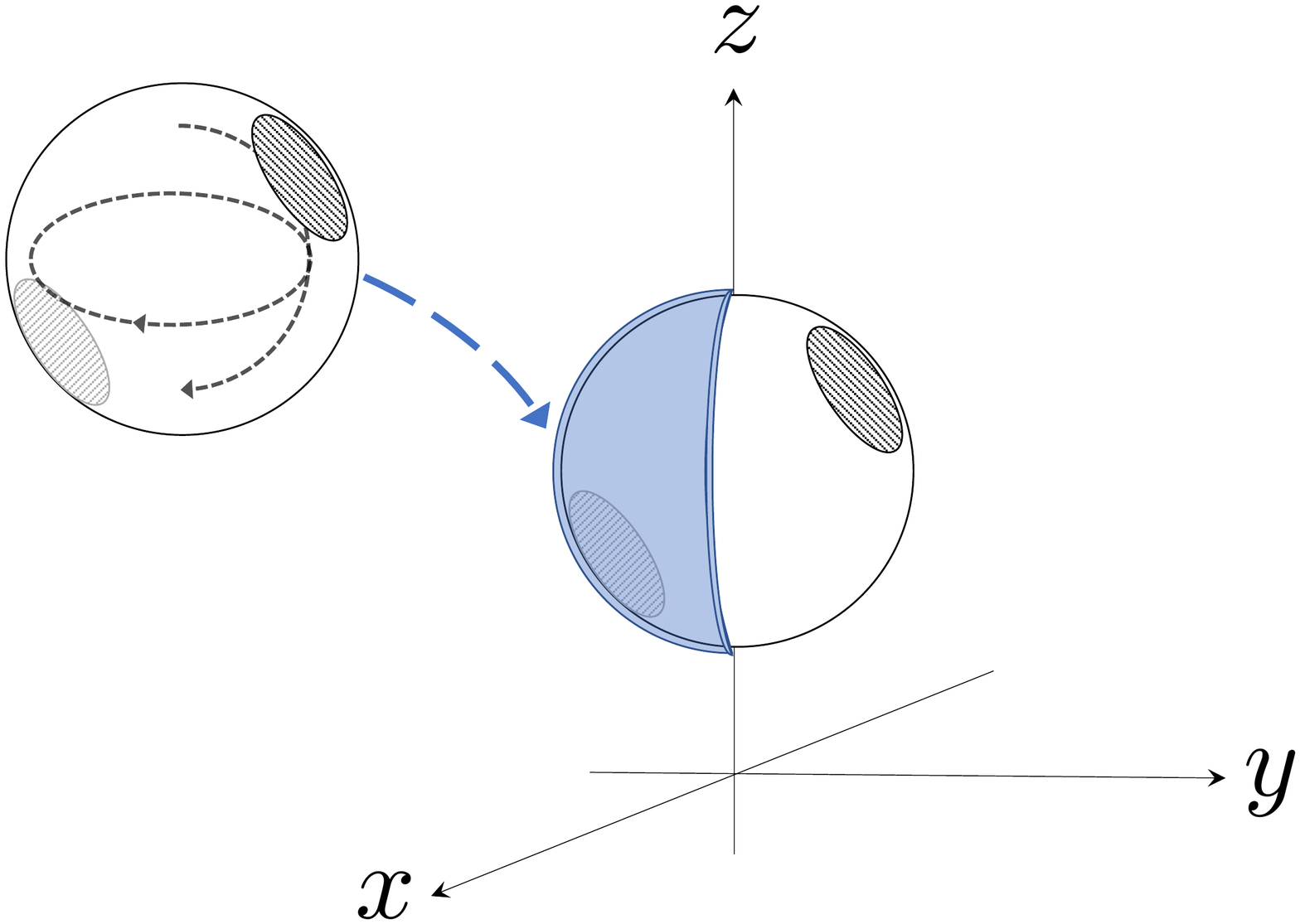}	\includegraphics[height=2.65cm]{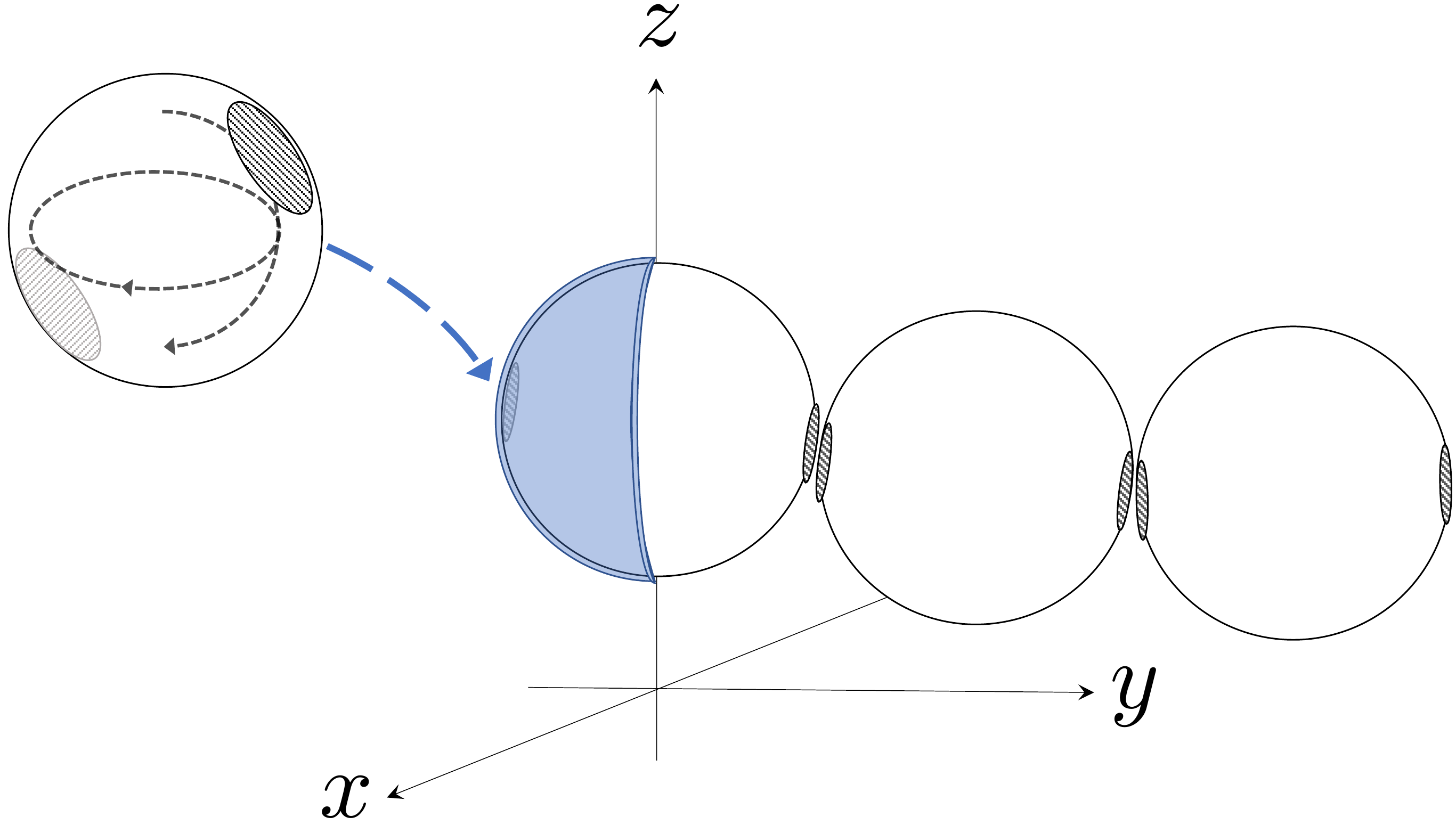}
	\caption{A schematic illustration of the calculation of $K_2$ and $K_4$ of particles under a gravitational field. The conformations of the $\gamma$-particle are sampled with MC allowing also translations again the gravitational field in the $z$-direction. } 
	\label{gravity_MC_sampling}
\end{figure}

	As a consequence, there may  now be  multiple reaction constants: 
	\begin{subequations}
	\begin{align}
		A_1 + A_1 &\stackrel{\mathrm{K_{2}}}{\rightleftharpoons}  A_2  \label{n2_reaction} \\
		&\dots  \notag \\
		A_1 + A_{l-2} &\stackrel{\mathrm{K_{k}}}{\rightleftharpoons}  A_{l-1} \label{reaction_l-2_l-1} \\
		A_1 + A_{l-1} &\stackrel{\mathrm{K}}{\rightleftharpoons}  A_l \label{reaction_l-1_l} \\
			&\dots  \notag \\
		A_1 + A_{n-1} &\stackrel{\mathrm{K}}{\rightleftharpoons}  A_n \label{nn_reactionn} 
	\end{align}
	\end{subequations}
	where the reaction constants $K_{2}\neq K_{3} \neq K_{k} \neq K $ may not be equal with each other (see Fig.~\ref{gravity_MC_sampling} for an illustration). Only beyond a certain chain length $k$ (Eq.~\ref{reaction_l-1_l}-\ref{nn_reactionn}), the equilibrium constant can be considered to settle. 

	The density $\rho$ is written as outlined in the previous section:
\begin{align}
	\rho &=\sum_{n=1}n \rho_n = \rho_1 +  2\rho_2 + 3 \rho_3 + ... +  n\rho_n  \notag\\
	&=  \rho_1 + 2 \rho_1^2K_{2} + 3 \rho_1^3K_{2} K_{3} + ...  + k\rho_1^kK_{2} ... K_{k}  + \notag \\ 
	 &+  (k+1)\rho_1^{k+1}K_{2} ... K_{k}K + ... + n  \rho_1^n(\prod_{m=2}^{k} K_m )K^{n-k}  \notag\\
	&= \rho_1 + \sum_{n=2}^{k} \left(n\rho_1^{n}\prod_{m=2}^{n} K_m \right) \notag\\ &+ \left(\prod_{m=2}^{k} K_m \right)\left(\sum_{n=k+1}n\rho_1^nK^{n-k}  \right) \label{density_rho1Ks}
\end{align}
The second sum in this expression converges to:
\begin{align}
\label{eq:geometric}
\sum_{n=k+1}n\rho_1^nK^{n-k} & = \rho_1^{k} \sum_{n=k+1}n(\rho_1K)^{n-k}  \notag \\
& = \rho_1^{k}\frac{ K\rho_1 (k+1-k K\rho_1)}{(\rho_1K-1)^2}
\end{align}
	
Similarly, the total number of clusters/chains in the system is:
\begin{align}
	\rho_c &=\sum_{n=1} \rho_n  
	= \rho_1 + \sum_{n=2}^{k}\left( \rho_1^{n} \prod_{m=2}^{n} K_m \right) \notag \\&+  \left(\prod_{m=2}^{k} K_m \right)\left(  \sum_{n=k+1}\rho_1^nK^{n-k} \right) 
\end{align}
where the infinite sum converges to:
\begin{align}
\sum_{n=k+1}\rho_1^nK^{n-k} = \rho_1^{k}\frac{K\rho_1}{1-K\rho_1}
\end{align}

Again, for given density $\rho$ and reaction constants $K_2$, $\dots$, $ K_k$ and $K$, the only unknown is the monomer fraction $X_1=\rho_1/\rho$ (Eq.~\ref{density_rho1Ks} and \ref{eq:geometric}). As Eq.~\ref{density_rho1Ks}  is a higher order polynomial, it should be (it is)  solved numerically.

	\section{Simulation Methods}

	\subsection{General patchy particle  pair potential}
\begin{figure}[t!]
			\centering
			\includegraphics[width=0.35 \textwidth]{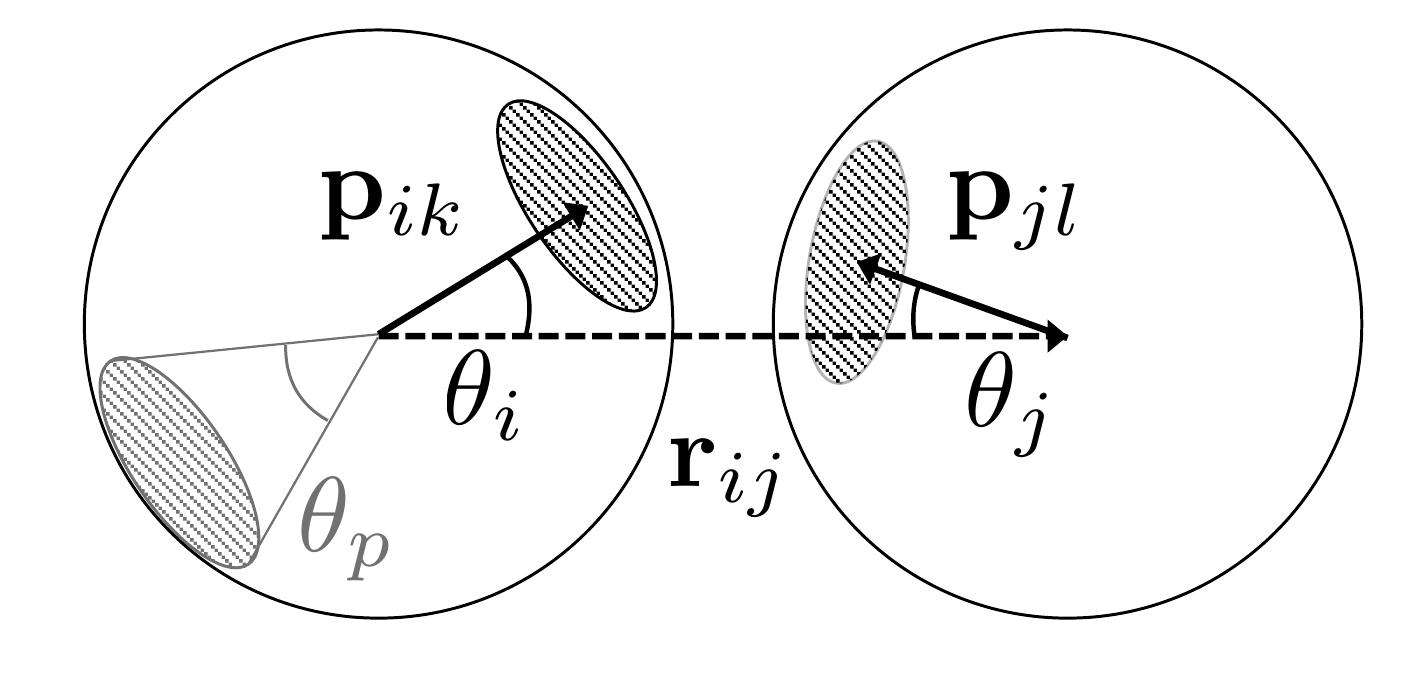}
			\caption{A schematic illustration of the interparticle vector $\mathbf{r}$ (dotted arrow), patch vectors $\mathbf{p}$ on each particle (solid arrows), and the angles $\theta$, and the patch size is defined by the angle $\theta_p$} 
			\label{patch_vecs}
			\end{figure}

	To test our extension of TPT1 under strong confinement, we simulate several systems with  a variety of potentials, from simple toy systems  to more accurate ones.
	The general expression for the  pair interaction $V_{\rm{pair}}$  between two  patchy particles $i$ and $j$ with orientation $\Omega_i$  and $\Omega_j$, respectively,  and  separated by a distance $r_{ij}$, is 
	\begin{align}
		V_{\rm{pair}}(r_{ij}, \Omega_i, \Omega_j) = V_{\rm rep}(r_{ij}) + \min_{1\le k,l \le n_p}   {V_{\mathbf{p}_{ik},\mathbf{p}_{jl}} (r_{ij}, \Omega_i,\Omega_j)}
		\label{eqVpair}
	\end{align}
	where $V_{\rm rep}$ denotes an isotropic repulsive potential and $V_{\mathbf{p}_{ik},\mathbf{p}_{jl}}$ patch-patch attractive interaction~\cite{Jonas2021}. The position of each patch in the particle reference frame is given by $n_p$ unit patch vectors  $\mathbf{p}$, which point from the particle's center to the center of the patch  (Fig.~\ref{patch_vecs}). The min function gives the minimum energy of the set of all possible patch-patch combinations and mimics the fact that we restrict our particles to form only one bond per particle pair. For our systems, the range and width of the patch interaction is relatively small, so that this condition is easily fulfilled. 
	
	The attractive patch–patch potential is defined as\cite{Jonas2021} 
	\begin{align}
          V_{\mathbf{p}_{ik},\mathbf{p}_{jl}} (r_{ij}, \Omega_i,\Omega_j) = V_{\rm attr}(r_{ij})S'(\Omega_i)S'(\Omega_j),
          \label{patchpotential}
	\end{align}
	where $V_{\rm attr}(r_{ij})$ is an isotropic attractive potential. As the patches turn away from each other, the patch-patch interaction becomes weaker as the area of overlap in between the patches decreases. This anisotropy of the patch interactions is captured by  the two switching functions $S'$ that are each a function of the orientation $\Omega$ of each particle.  See  Ref.~\onlinecite{Jonas2021} for more details.

	\subsubsection{Square well pair potential}
\begin{figure}[t!]
	\centering
	\includegraphics[width=0.45\textwidth]{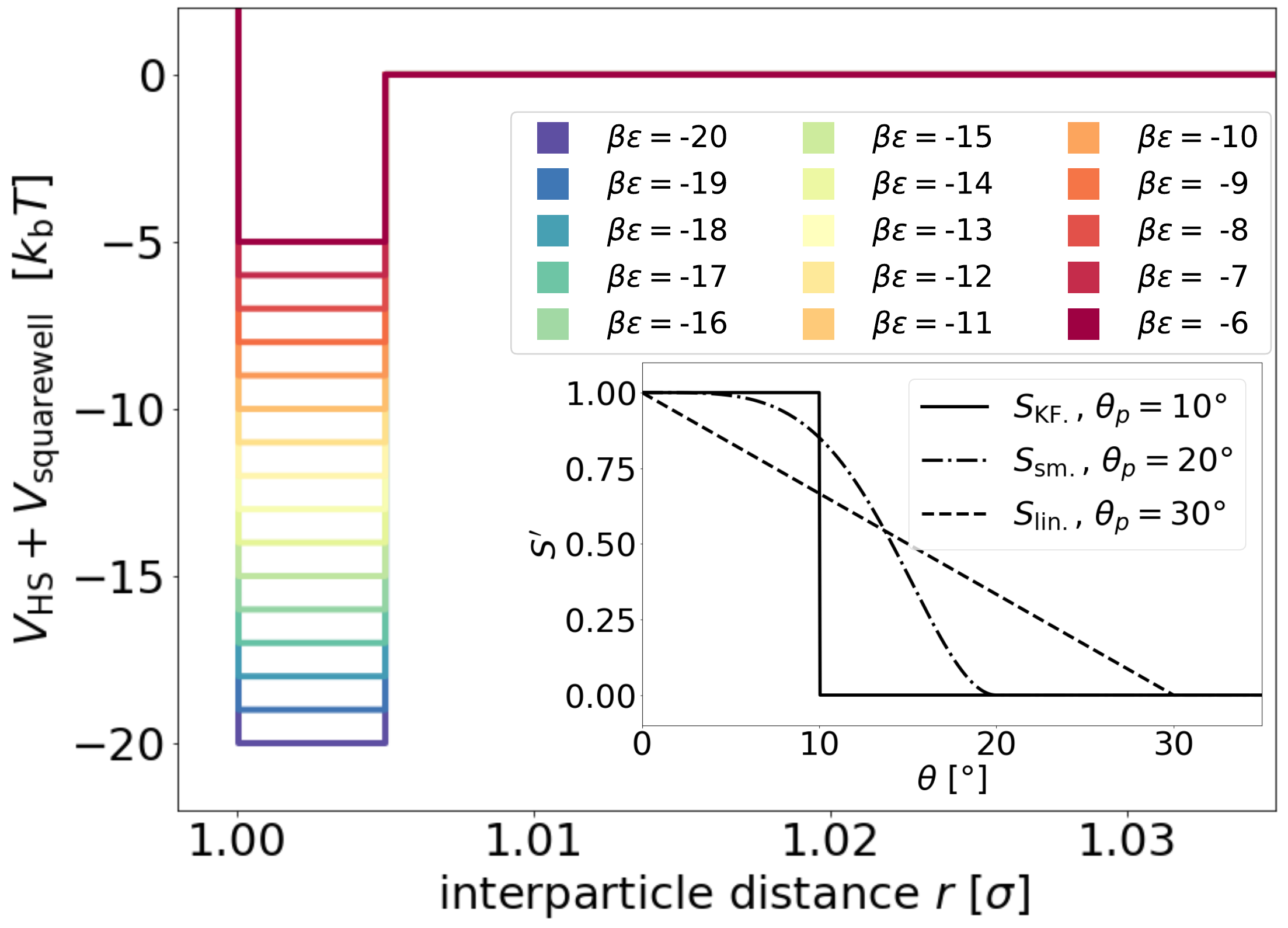}
	\caption{ The toy model radial potential composed of the hard sphere repulsion $V_{\rm HS}$ and square well attraction $V_{\rm square well}$. The inset shows the three switching functions $S_{\rm KF}$ (conical, Eq.~\ref{S_KF}), $S_{\rm sm.}$ (smooth, Eq.~\ref{S_smooth}), and $S_{\rm lin.}$ (linear, Eq.~\ref{S_linear}).  } 
	\label{sqw}
\end{figure}

	The simple toy systems employs a  hard sphere  with diameter $\sigma$  and a square well attraction (Fig.~\ref{sqw}). The square well  potential  is 
\begin{align}
	&V_{\rm rep}(r)=
	\begin{cases}
		\infty &r<\sigma\\
		0 & \rm otherwise
	\end{cases} 
\end{align}
 together with  a square well attraction:
	\begin{align}
		&V_{\rm attr}(r)=
		\begin{cases}
			\beta\epsilon &\sigma<r\leq\sigma + \delta\\
			0 & \rm otherwise
		\end{cases} 
	\end{align}
	where $\delta=0.005\sigma$, $\beta=1/k_{\rm B}T$ the inverse temperature with $k_{\rm B}$ the Boltzmann constant, $\beta\epsilon\in[-20,-5]$ corresponding to a reduced temperature $T^*=-1/\beta\epsilon \in[0.05,0.20]$.  Note that  this square well is rather narrow.
	
	The  switching function $S'(\Omega)$ is defined as:
	\begin{itemize}
		\item  conical, also known as the Kern-Frenkel potential\cite{Bol1982,Kern2003}:
		\begin{align}
			&S_{\rm KF}(\theta,\theta_{\rm p})=
			\begin{cases}
				1&  \theta\leq\theta_{p} \\ 
				0 &\theta>\theta_{p}
				\label{S_KF}\end{cases} 
		\end{align}
		\item smooth \cite{Guo2015}:
		\begin{align}
			&S_{\rm sm.}(\theta,\theta_{\rm p})=
			\begin{cases}
				\frac{1}{2} (1-\cos(\pi \frac{\cos(\theta) - \cos(\theta_{\rm p})}{1- \cos(\theta_{\rm p})}))&  \theta\leq\theta_{p}\\ 
				0 &\theta>\theta_{p}
				\label{S_smooth}\end{cases} 
		\end{align}
		\item or linear:
		\begin{align}
			&S_{\rm lin.}(\theta,\theta_{\rm p})=
			\begin{cases}
				1 - \theta/\theta_{\rm p}&  \theta\leq\theta_{p}\\ 
				0 &\theta>\theta_{p}
				\label{S_linear}\end{cases} 
		\end{align}
	\end{itemize}
	where $\theta$ is the angle between the interparticle vector $\mathbf{r}$ and the patch vector $\mathbf{p}$, and $\theta_{\rm p}$ is the patch size  (Fig.~\ref{patch_vecs}). As shown in the inset in Fig.~\ref{sqw}, the patch size $\theta_{\rm p}$ was varied from 10, 20, and 30$\degree$ for the $S_{\rm KF}$, $S_{\rm sm.}$ and $S_{\rm lin.}$ switch functions, respectively. 
	
	\subsubsection{The critical Casimir pair potential}

\begin{figure}[t!]
	\centering
	\includegraphics[width=0.48\textwidth]{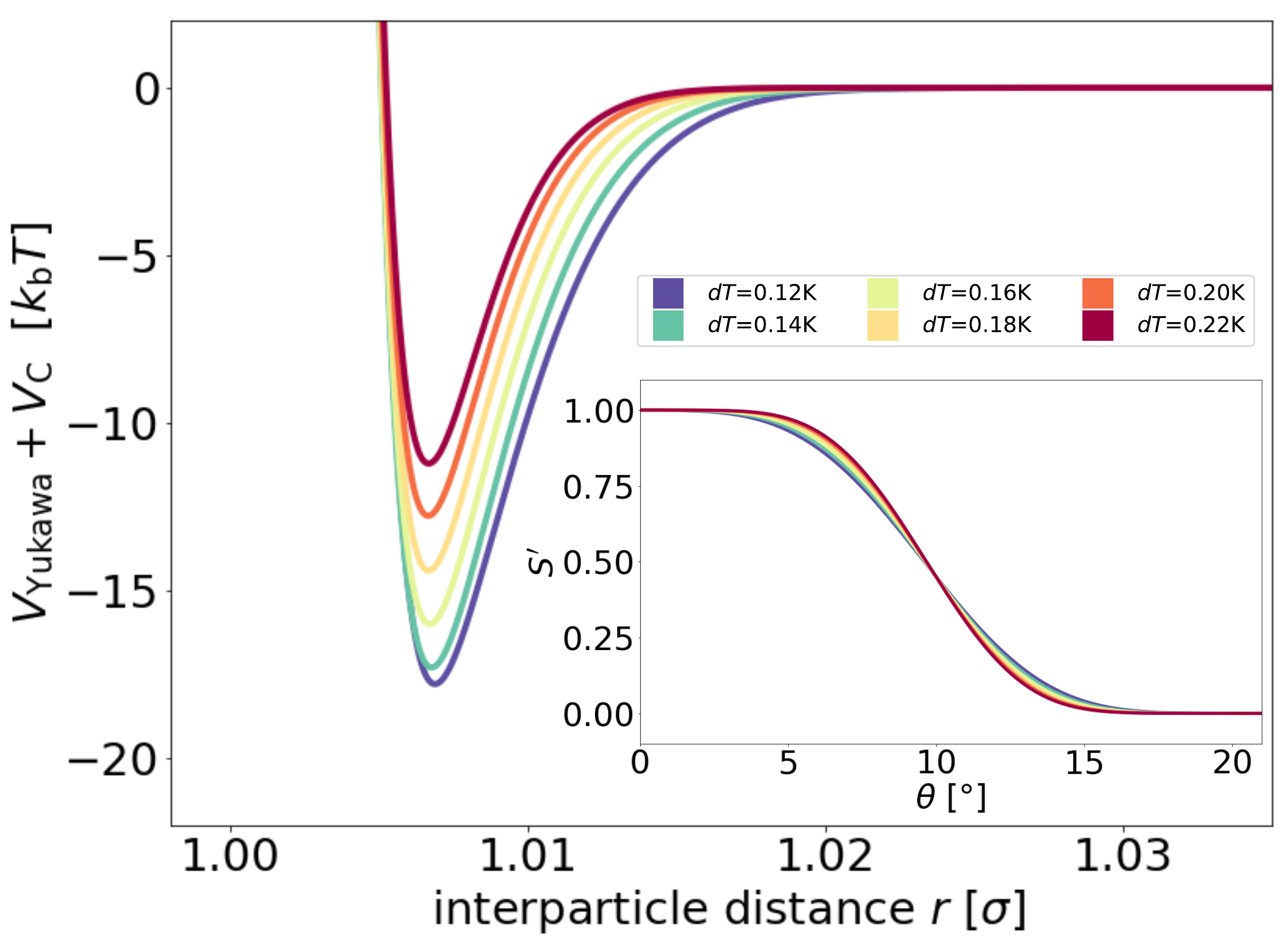}
	\caption{The patchy particle radial potential for dipatch particles composed of Yukawa repulsion $V_{\rm Yukawa}$ (Eq.~\ref{V_yukawa}) and  critical Casimir attraction $V_{\rm C}$ (Eq.~\ref{V_C}). The potential uses variables $w$=0.462, $\Upsilon$=-0.09$e/$nm$^2$, and $\theta_p^{\rm eff}$=19.5$\degree$. The inset shows the switching functions that are additionally a function of $dT$.  } 
	\label{casimir}
\end{figure}

	The accurate effective critical Casimir potential of the dipatch particles has its own radial dependence and switching function. In previous work, we optimized the potential (Fig.~\ref{casimir}), based on physical dimensions of the dipatch particles and theoretical critical Casimir potentials,  to reproduce chain length distributions and bending rigidities observed in experiment over a weak to strong interaction range as function of the temperature\cite{Jonas2021}.
	
	The repulsive potential is given by the electrostatic Yukawa potential
	\begin{align} 
	V_{\rm rep} (r) = &V_{\rm Yukawa}(r) \notag  \\
	= & \begin{cases}
		\infty,  &  r\leq \sigma\\ 
		U_0 \exp(-\kappa (r-\sigma)),&r>\sigma
	\end{cases},\label{V_yukawa}
	\end{align}
	with
	\begin{equation}
		U_0 = \frac{Z^2\lambda_{\rm B}}{(1+\kappa \sigma_c/2)^2 r}, \label{yukawaU0}  
	\end{equation}
	where $Z=\pi\sigma_c^2\Upsilon$ is the charge of the particles,  $\sigma_c$ the diameter of the charged colloids, $\Upsilon$ the surface charge density, $\lambda_{\rm B}=\beta e^2/4\pi\epsilon$ the Bjerrum length of the solvent with $\epsilon$ is the permittivity of the solvent, $e$ the elementary charge. The screening length, i.e. Debye length, is defined as $\kappa^{-1}=\sqrt{\epsilon k_{\rm B}T/e^2\sum_{i}\rho_i}$ where $\rho_i$ is the number density of monovalent ions in the solvent\cite{AdrianParsegian2005}.  
	The model uses  $\sigma_c$=2.0$\mu$m, $\lambda_{\rm B}$=2.14nm, $\kappa^{-1}$= 2.78nm and the surface charge density was optimized to $\Upsilon$=-0.09$e/$nm$^2$. See Ref. \onlinecite{Jonas2021} for more details.
	
        The attractive potential is based on the (isotropic) critical Casimir potential $V_C$ between two spherical particles:
		\begin{equation}
		V_C(r) = - \frac{A}{B} \exp \left(-\left(\frac{r-\sigma}{B}\right)^2\right) \label{V_C}
	\end{equation}
	where $A$ and $B$ are fit parameters depending on the wetting set to $w=0.462$ and temperature $dT\in[0.12,0.22]K$ and can be found in Appendix B1 in Ref.~\onlinecite{Jonas2021}.  The wetting $w$ is determined by the interplay between the patch material and the surrounding binary liquid and the temperature is defined as its distance from the phase separation temperature $T_{cx}$ of the binary liquid $dT=T-T_{cx}$.
	
	The switching function was obtained by explicit integration using $ V_{\rm Yukawa}$ and $V_C$ at an effective patch width of $\theta_{\rm p}^{\rm eff}=19.5\degree$ and fitted to the following functional form: 
		\begin{equation}
		S'(\theta)= \exp\left(\sum_{l=2}^{8} c_l\theta^l\right),
\end{equation}	
where the coefficients $c_l$ are given in Table VI. in Ref. \onlinecite{Jonas2021}.

	\subsection{External potential}
	The external potential  $V_{\rm ext}$ at quasi-2D confinement prohibits the translation in the $z$-direction via: 
	\begin{align}
		&V_{\rm quasi-2D}(z)=
		\begin{cases}
			0 &  z=0 \\ 
		 \infty &z\neq0
		\end{cases} 
	\end{align}
	which mimics two hard walls separated by the particle diameter. 
	
	The external gravitational potential is composed of two terms: a hard wall represented by a steep Lennard-Jones  potential $V_{LJ}$ and a gravitational potential $V_g$ that  depends, among other things, on the mass of the particle. See the appendix B2 of Ref.~\onlinecite{Jonas2021} for more details. 
	The resulting total gravitational  potential is:
	\begin{align}
		&V_{\rm gravity}(z)=
		\begin{cases}
			V_{\rm LJ}(z) &  z\leq z_{\rm cut} \\ 
			V_g(z) &z>z_{\rm cut}
		\end{cases} \\
		&=
		\begin{cases}
			4\epsilon_{ \rm LJ}\left(\left( \frac{\sigma}{z}\right) ^{12}-\left( \frac{\sigma}{z}\right) ^{6} + \frac{1}{4}\right), &  z\leq z_{\rm cut} \\ 
			-F_gz -b, &z>z_{\rm cut}
		\end{cases} \label{Vgravity}
	\end{align}
	where  $\epsilon_{\rm LJ}$ is an arbitrary (high) value set to 500$k_{\rm B}T$ and the gravitational force $F_g$ is varied from -3.85, to -7.70, and to -11.55 $k_{\rm B}T/\sigma$.  
	These gravitational forces ranges from 0.5, 1.0, and 1.5 times the gravitational force of the dipatch particle of interest, respectively.  Parameters $b$ and $z_{\rm cut}$ are chosen such that both the potential and the force are  continuous at $z_{\rm cut}$.
	
		\subsection{The system's potential energy}
	 	The system's potential energy is a sum over all pair potentials and the external field
	\begin{equation}
		V = \sum_{i<j}^N V_{\rm pair}(r_{ij},\Omega_i,\Omega_j) + \sum_{i}^NV_{\rm ext}(z_i)
	\end{equation}
	where $i$ and $j$ run over the $N$ colloidal particles. The external potential $V_{\rm ext}$ is either $V_{\rm quasi-2D}$ or $V_{\rm gravity}$.
 
	\subsection{The calculation of $\Delta$ \label{sec:Delta_calculation}}

	The integral in Eq.~\ref{Delta} is performed using Monte Carlo integration via:
	\begin{align}
					\Delta  = V \left< g(\mathbf{r})  \left\langle f(r,\Omega_{\alpha},\Omega_{\gamma})\right\rangle _{\Omega_{\alpha},\Omega_{\gamma}} \right>_{V} \label{Delta_Vavg}   
				\end{align}
	where the average over the radial distribution function and Mayer function takes place in the integration volume $V= \int d\mathbf{r}$.  The monomers are indicated by $\alpha$ and the other reactant is the $\gamma$-particle. Each reaction constant $K_m$ in the polymerization is thus defined by the orientational and positional distribution of the reactants and  a separate computation  of the corresponding $\Delta_m$ must be done.  
	
	Correct determination of the averages is key to calculate $\Delta$. There are two options to measure the average: (1) sample homogeneously over space multiplied with the probability distribution 
	or (2) sample from the correct  distribution. This applies to both orientational $\left\langle \right\rangle_{\Omega_{\alpha},\Omega_{\gamma}}$ as well as the translational $\left\langle\right\rangle_{V}$ parts of the integration. An advantage of Monte Carlo integration is that it is capable of evaluating both averages simultaneously. 

	The $\alpha$-particle, and the $\gamma$-particle if it is unbound, are free monomers and their orientational distribution is uniform. Giving them random orientations samples the distribution as option (2). The  $\gamma$-particle's $z$-positional distribution is described by the (Boltzmann distribution of the) external potential $V_{\rm ext}$, and is sampled as option (2). 
	If the $\gamma$-particle is restricted to bound configurations, its orientational and positional distributions are additionally described by the pair potential.  So instead of uniformly sampling orientational and positional space, its configurations are sampled from a chain with MC (Fig.~\ref{q2D_MC_sampling}b).  Note that this MC sampling of the $\gamma$-particle's bonding configurations  is independent of the MC sampling of $\Delta$ (Eq.~\ref{Delta_Vavg}) and may be performed on-the-fly or  beforehand. Thus, the contribution of the orientations $\Omega_{\alpha}$ and  $\Omega_{\gamma}$ on the average is incorporated using option (2).
	
	The contribution of the inter-particle distance $r$ on the average is sampled by placing the $\alpha$-particle randomly in a hemispherical shell  of volume $V$ around the $\gamma$-particle.  The radial distribution function then gives the probability of finding the $\alpha$-particle at distance $r$. Thus, the contribution of the inter-particle  distance  on the average is incorporated using option (1).
	
	\subsection{Simulation Details}
	\subsubsection{Explicit MC sampling of chain length distributions}
	Systems with the square well radial potential were simulated  with MC for $N=$1000 divalent particles in a square box with periodic boundary conditions of length  51.17$\sigma$ or 62.67$\sigma$ to resemble a density of  $\rho=0.382$ or $0.255$  $N/\sigma^2$, respectively.  The densities were chosen to yield  reasonable  chain length distributions, i.e. that probabilities for longer chain were nonzero.
	The systems with the critical Casimir potential were taken from Ref.~\onlinecite{Stuij2021}. These simulations were done in a rectangular box with periodic boundary conditions of dimensions $43.5\sigma\times60\sigma\times43.5\sigma$ with 666, or 1000 
	particles corresponding to  $\rho=0.255$ or $0.382$ $N/\sigma^2$, respectively.
	
	Starting from a random  starting configuration MC moves were performed to equilibrate and measure the systems as explained in detail in Ref.~\onlinecite{Jonas2021}. Depending on the  interaction strength and form of switch function, the equilibration consisted of $1\times 10^4$ to $6\times 10^{4}$ MC cycles and the measurements of  $5\times 10^4$ to $2\times10^5$ MC cycles. Each MC cycle consists of $5\times 10^{5}$ single particle (95\%) and cluster moves (5\%).
	
	The chain length distribution was measured from three independent simulations after each MC cycle by counting the number $N_n$ of chains of length $n$ and normalize by the total number of chains in the system yielding $P_n= N_n/\sum_{n=1}N_n$. Irrespective of the use of the discontinuous (square well) or continuous  (critical Casimir) attractive potential, a bond is defined for a pair of particles if $V_{\mathbf{p}_{ik},\mathbf{p}_{jl}} (r_{ij}, \Omega_i,\Omega_j)<0 k_{\rm B}T$.

	\subsubsection{MC sampling of $\Delta$}
	For the calculation of the volume average $\left<\right>_V$ in Eq.~\ref{Delta_Vavg} we employ three loops. 
	In the first loop of $10^4$ cycles, the configurations of the $\gamma$-particle are sampled. For the quasi-2D system, a monomer and dimer configuration are sufficient as there are only two reaction constants. For the gravitational systems,  equilibrated chains with length $l\leq15$ are decorrelated with $10^5$ single particle MC moves. The position and orientation of the hemi-sphere of the free site is saved as the $\gamma$-particle.
	In the second loop of $10^2$ cycles, the new position of the $\alpha$-particle is set to $\vec{r}_{\alpha}=\vec{r}_{\gamma} + r\vec{e}$ where $r$ is a random distance $r \in [\sigma,\sigma+\delta]$ and $\vec{e}$ a random unit vector pointing to the $\gamma$ hemi-sphere. The  RDF gives the probability of finding the particles at positions $\vec{r}_{\alpha}$ and $\vec{r}_{\gamma}$  in the hard particle reference, more details are in section \ref{sec:rdfs}. 
	In the third loop of $10^3$ cycles, the $\alpha$-particle is given random orientations and the Mayer function is calculated for the sampling of the $\left<\right>_V$.	To improve the calculation of the average, the $\alpha$-particle was given six sites instead of two. In total there are $2 \times 6$ possible patch combinations, two from the $\gamma$ and six from the $\alpha$-particle,  leading to a simple correction of $1/12$.
	The final step to calculate $\Delta$ is to multiply the $\left<\right>_V$  with the volume of the hemispherical-shell $V$.  
	The loops may be repeated one to three times independently, depending on the convergence. 

	\subsubsection{The radial distribution function \label{sec:rdfs}}
	The radial distribution function (RDF) of the hard sphere reference fluid is important for the radial component of $\Delta$.  The quasi-2D system is radially isotropic in the $(x,y)$-plane, we thus may use a heuristic RDF $g_{\rm HD}(r)$ of hard disks \cite{Santos2016a,Yuste1993a}.
	For the gravitational confined systems, MC simulations are performed to measure the radial distribution of hard spheres at various densities and gravitational fields. 
	The RDF was saved to a file and uses the distance between the particles $r=|\vec{r_{12}}|$, and the z-coordinates of the particles  $z_1$ and $z_2$ as variables. A bin width of $dr=0.05\sigma$ and $dz=0.01\sigma$ in combination with a simple flooring of the bin was sufficient to determine the corresponding RDF during the $\Delta$-calculation.
	
	The reference hard particle diameter $d$ when using the square well potential is simply $d=\sigma$ as the attractive part of the potential is taken as the perturbation. While for the critical Casimir patchy particle potential, we use the WCA separation to determine $d$ which, in principle, is also a function of the orientation, as $V_{\rm pair}$ is a function of $r$, $\Omega_i$, and $\Omega_j$ (Eq.~\ref{eqVpair}) \cite{Weeks1971,HANSEN2013149}. However,  the attractive potential is rather narrow and to avoid a variable reference diameter, $d$ is calculated at maximum attraction, i.e. $V_{\rm pair} =   V_{\rm Yukawa}(r) + V_C(r)$,  and used as a constant.

	\subsubsection{Numerically solving $X_1$}
	Given the density and the reaction constants, we determine the monomer fraction $X_1$ which is bound to $X_1\in [0,1]$  using the \textit{solve}-function from the \textit{SymPy}-Python package\cite{10.7717/peerj-cs.103}. There may be multiple (imaginary) solutions to the equations \ref{X1_KK2} or \ref{density_rho1Ks}, and solutions containing a very small imaginary part ($\textrm{Im}({X_1})<10^{-30}$) were accepted. 
	 In case of multiple solutions for $X_1$,  the smallest one was taken to solve the chain length distribution.

		\begin{figure}[t!]
			\centering
			\includegraphics[width=0.48\textwidth]{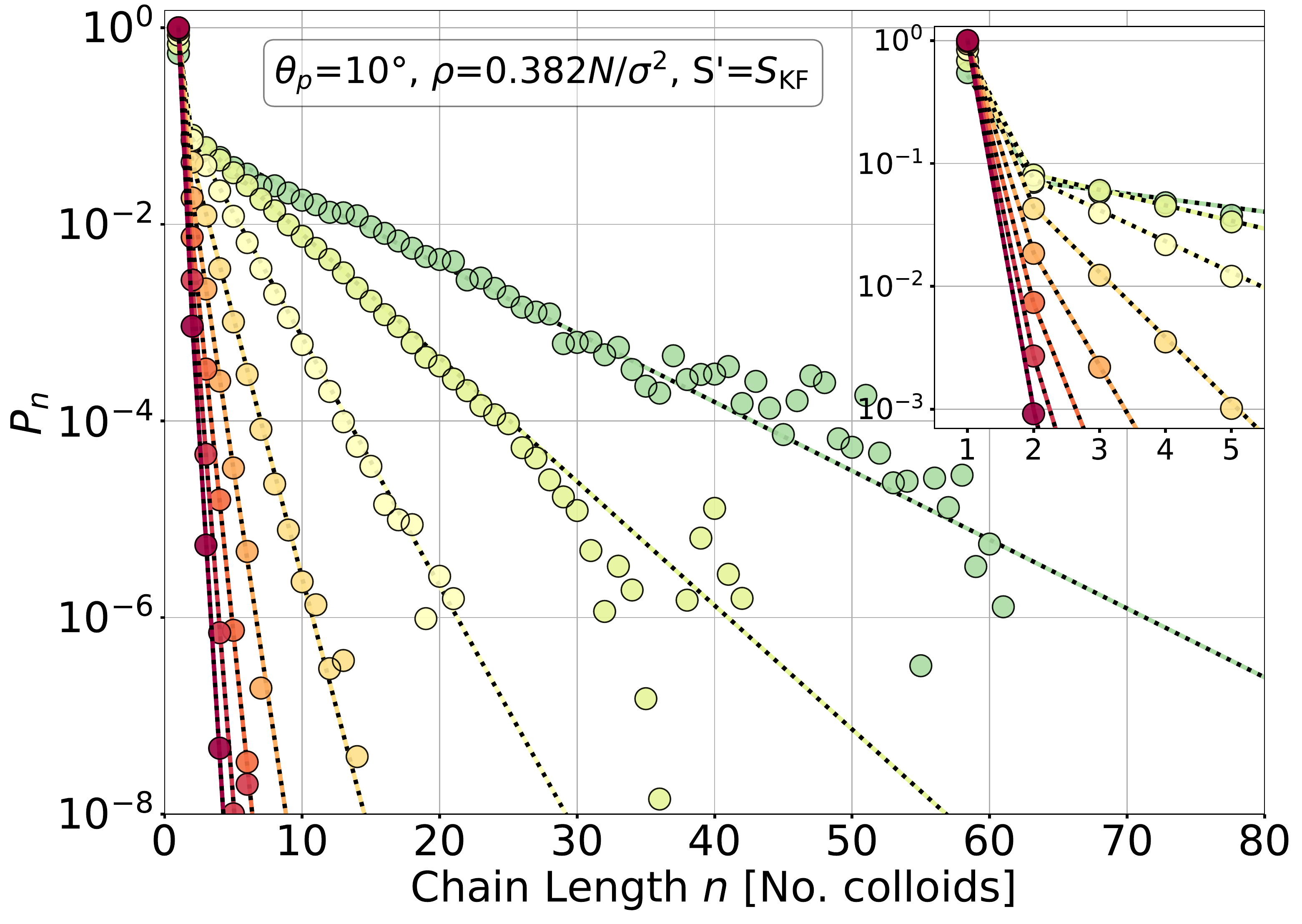}
			\includegraphics[width=0.48\textwidth]{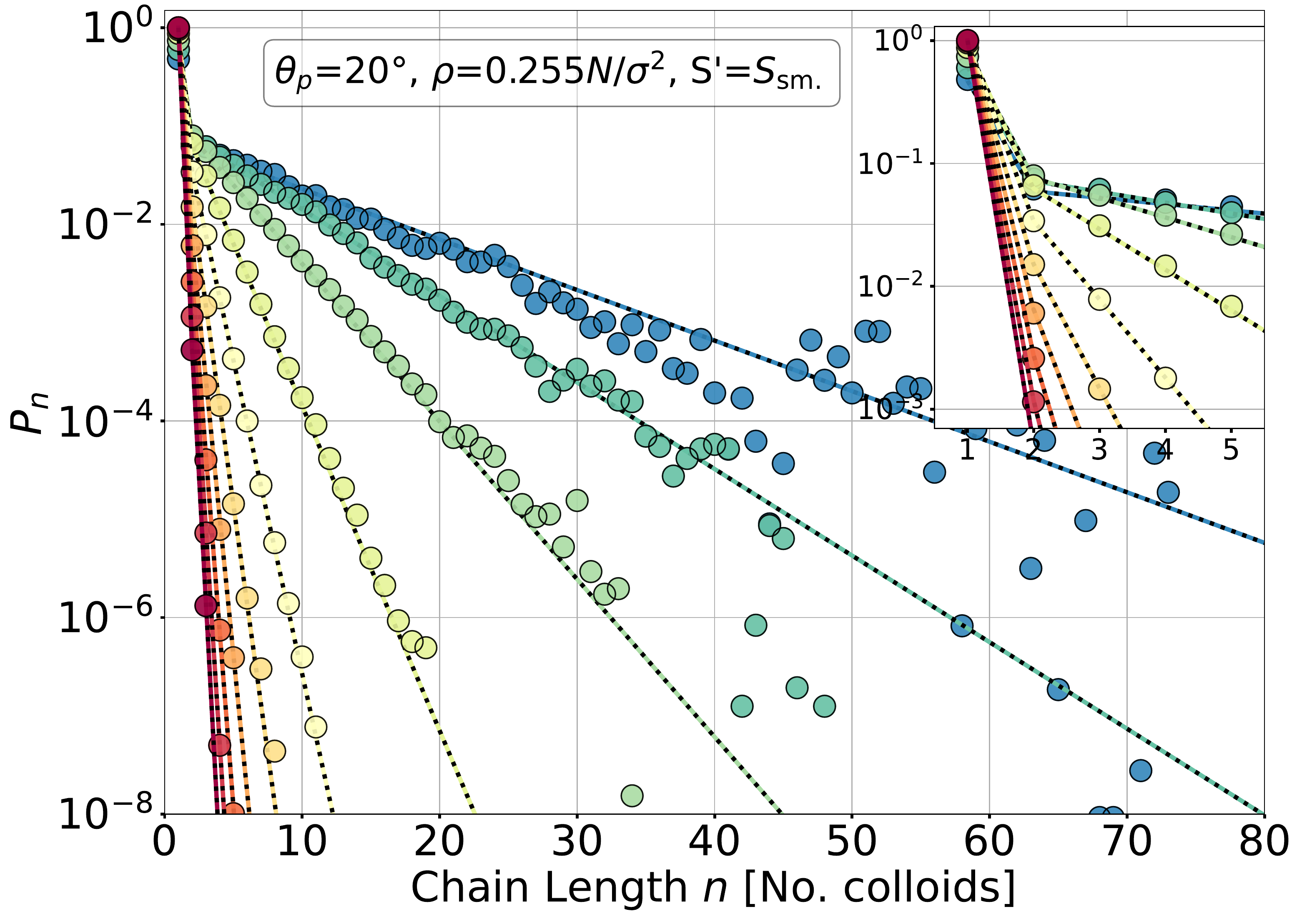}	
			\includegraphics[width=0.48\textwidth]{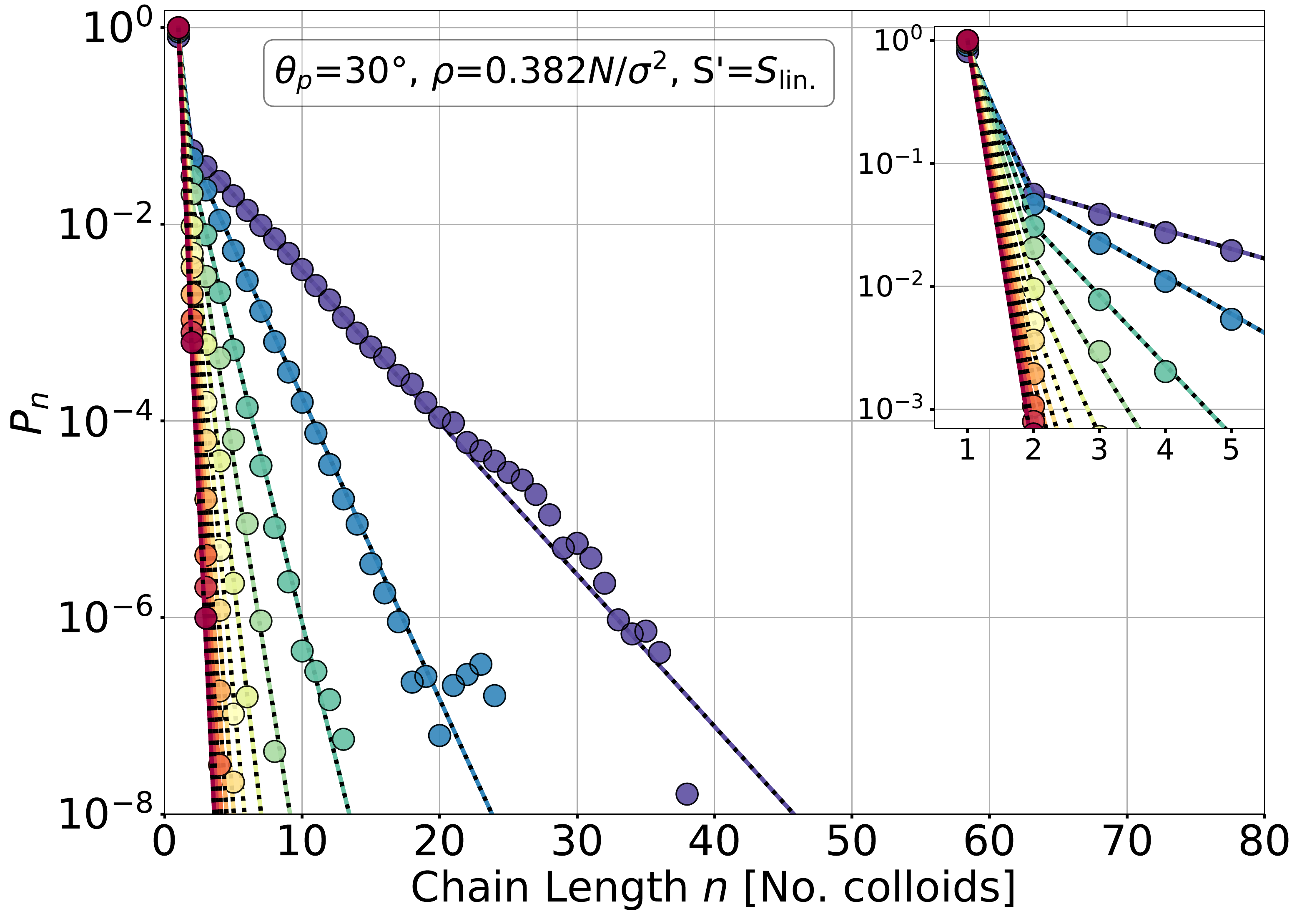}
			\includegraphics[width=0.48\textwidth]{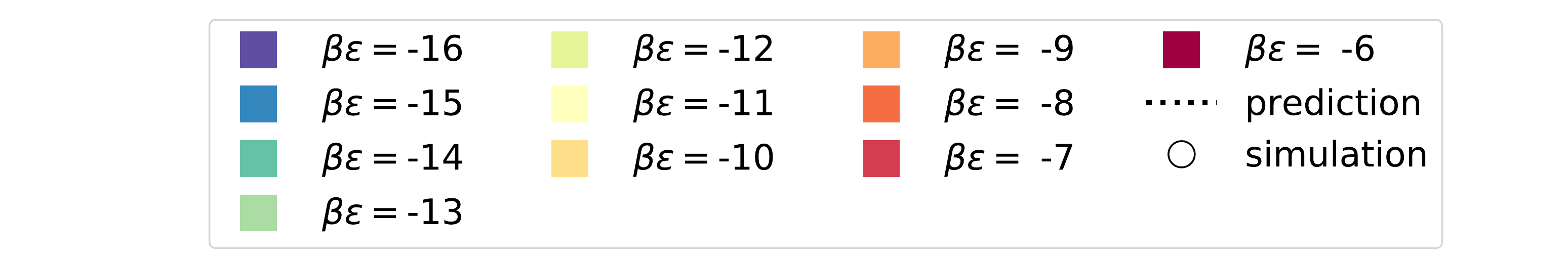}		
			\caption{Predicted (solid lines) and simulated  (dots) chain length distributions of systems  at various interactions strength ($\beta\epsilon$ indicated by colors), switch function type $S$ defined in Eq.~\ref{S_KF}-\ref{S_linear}, and density $\rho$ [$N/\sigma^2$]. The inset shows the probabilities at small chain lengths. } 
			\label{CLD_prediction}
		\end{figure}
		
	\section{Results and Discussion }

		To check the predictions of the extended Wertheim theory, we compare predicted and simulated chain length distributions in quasi-2D  of systems with a square well radial potential for the three different switching function at a wide range of attractive strengths.
		The resulting  simulated (dots) and predicted (solid lines) chain length distribution are shown in Fig.~\ref{CLD_prediction}. The top panel shows dipatch particles with a conical switch function $S'=S_{\rm KF}$ with patch size $\theta_{\rm p}$=10$\degree$ at $\rho=0.382N/\sigma^2$, the middle panel $S'=S_{\rm sm.}$ with $\theta_{\rm p}$=20$\degree$ at $\rho=0.255N/\sigma^2$, and the lower panel a $S'=S_{\rm lin.}$ with $\theta_{\rm p}$=30$\degree$ at $\rho=0.382N/\sigma^2$.
		The hallmark of all distribution is a clear increased monomer density that does not follow the exponential decay of the sequential polymerization, and is well described by the theory.
		All systems show excellent agreement between model predictions and simulations for all  radial interaction strengths, forms of switching function and densities.
			
		These quasi-2D systems allow for determination of the associated excess rotational entropy which can be directly  determined from the ratio of chain probabilities.
		 For the initial dimerization, the two monomers lose an excess rotational entropy $2\times F_{\rm rot}$ and gain a bonding free energy $F_{\rm bond}$, making $\exp( -\beta ( F_{\rm bond} - 2F_{\rm rot}  )) = {P_2}/{P_1} = K_2 \rho_1$.  
		 For the subsequent polymerization steps, only one monomer loses $F_{\rm rot}$ and $F_{\rm bond}$  is gained, thus $\exp(-\beta (F_{\rm bond} -  F_{\rm rot} ))= {P_3}/{P_2} = K \rho_1$.  

		\begin{figure}[t!]
	\centering
	\includegraphics[width=0.48\textwidth]{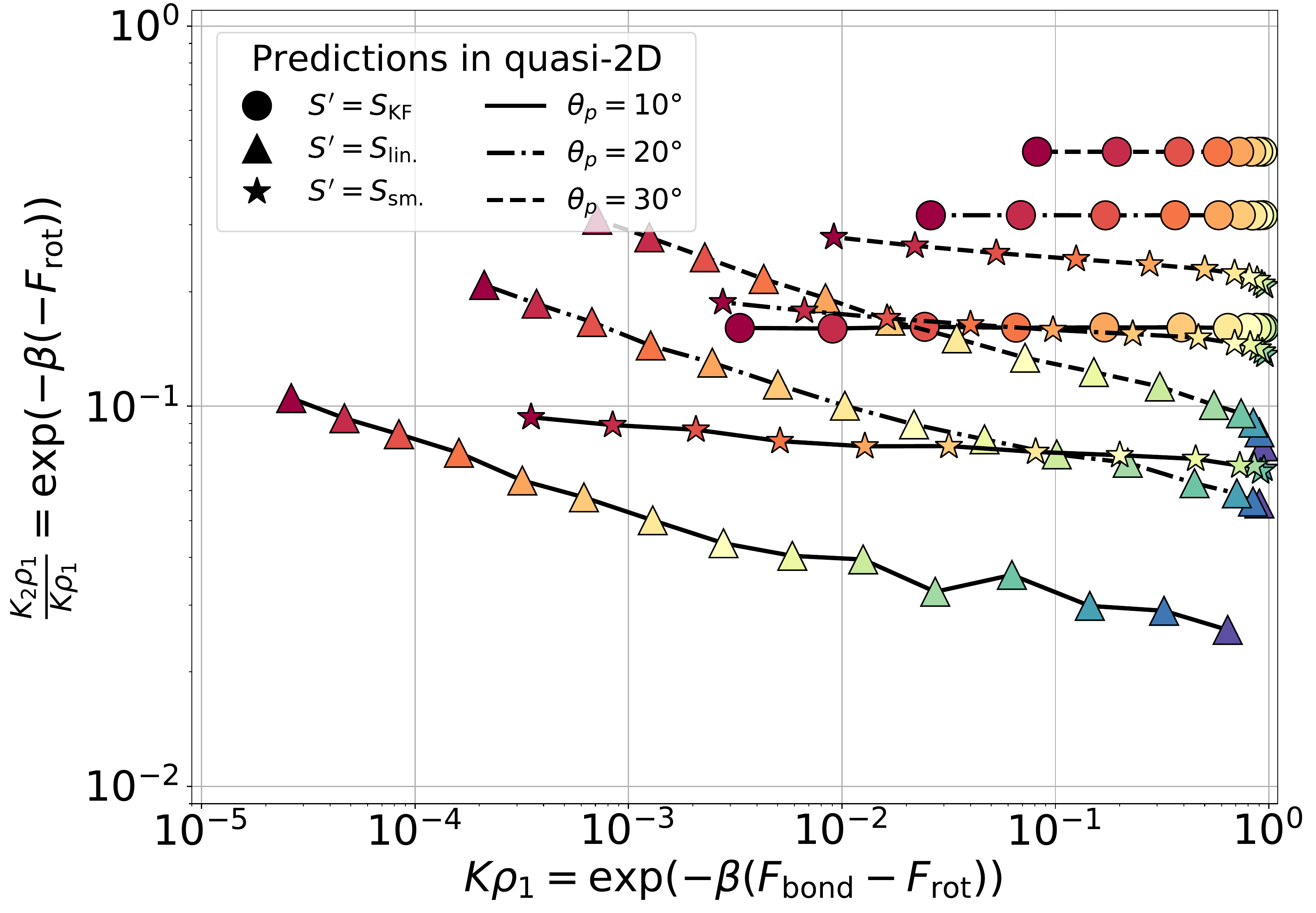}
	\includegraphics[width=0.48\textwidth]{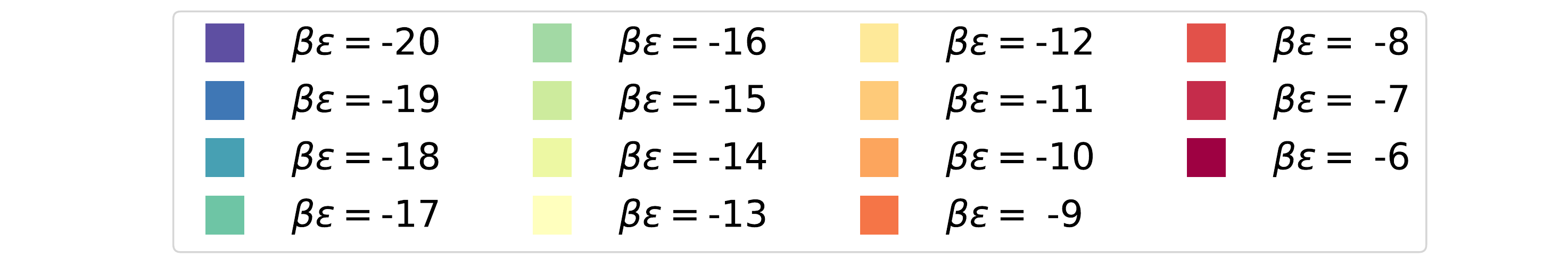}
	\caption{The predicted (exponents of the) excess rotational entropy $\beta F_{\rm rot}$ and the bonding free energy $\beta F_{\rm bond}$ of square well radial potentials with various switching functions in quasi-2D confined  at $\rho=0.382N/\sigma^2$. } 
	\label{rot_excess}
\end{figure}

		In  Fig.~\ref{rot_excess},  the resulting (exponents) of these free energies are collected for various patch types and sizes, and interaction strength at $\rho=0.382N/\sigma^2$.  For the conical potential  with $S'=S_{\rm KF}$, the excess rotational entropy is independent of the interaction strength and only a function of the patch size $\theta$. In that case, the bonding probability  is a purely geometrical factor and, as expected, larger patch sizes contain less excess rotational entropy. 
		Non-conical switch functions, namely $S_{\rm sm.}$ and $S_{\rm lin.}$, are dependent on the interaction strength. Here, stronger interaction strengths lead to stiffer chains, and thus to more excess rotational entropy for the monomers. 
		
	\begin{figure}[t!]
		\centering
		\includegraphics[width=0.48\textwidth]{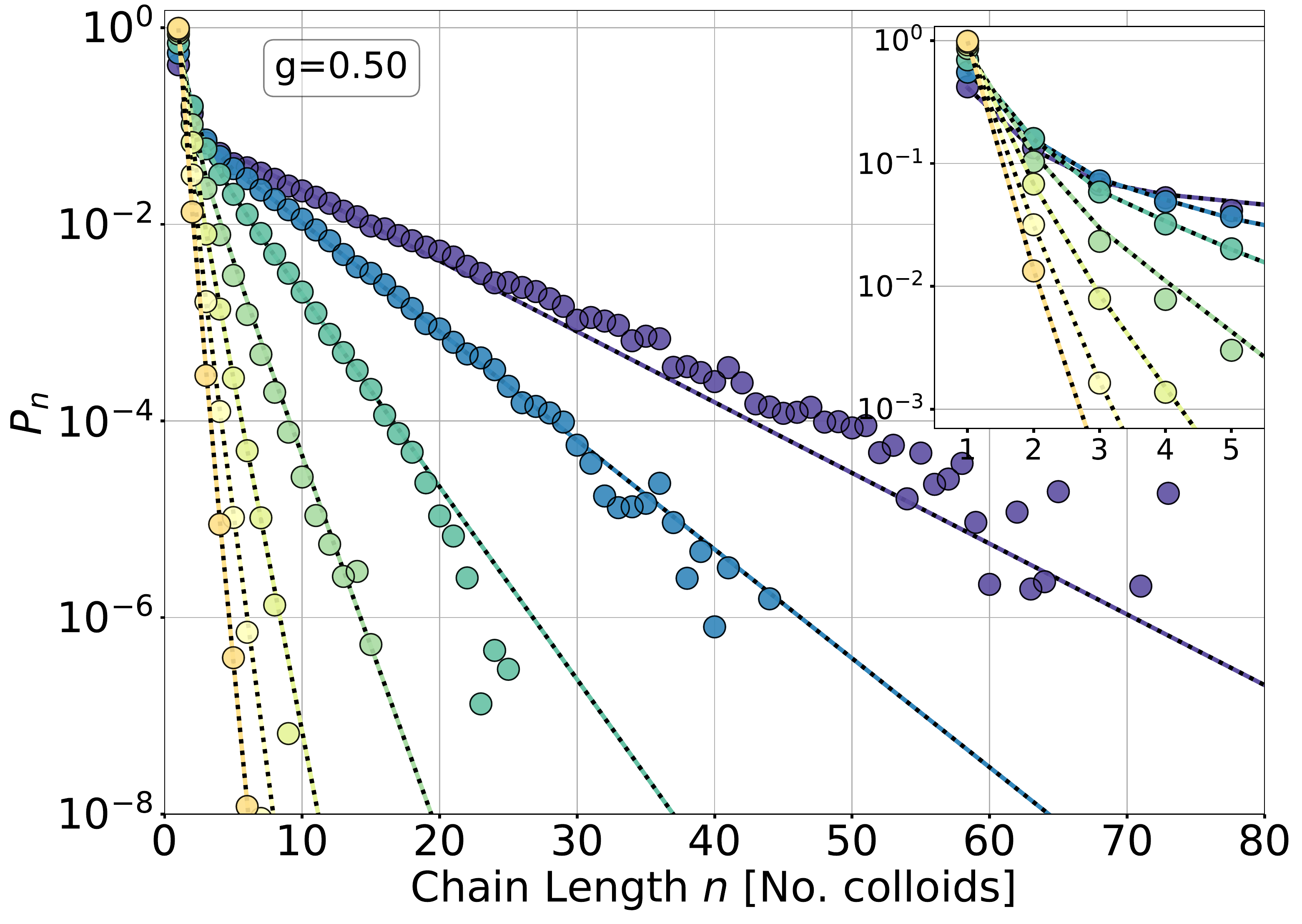}
				\includegraphics[width=0.48\textwidth]{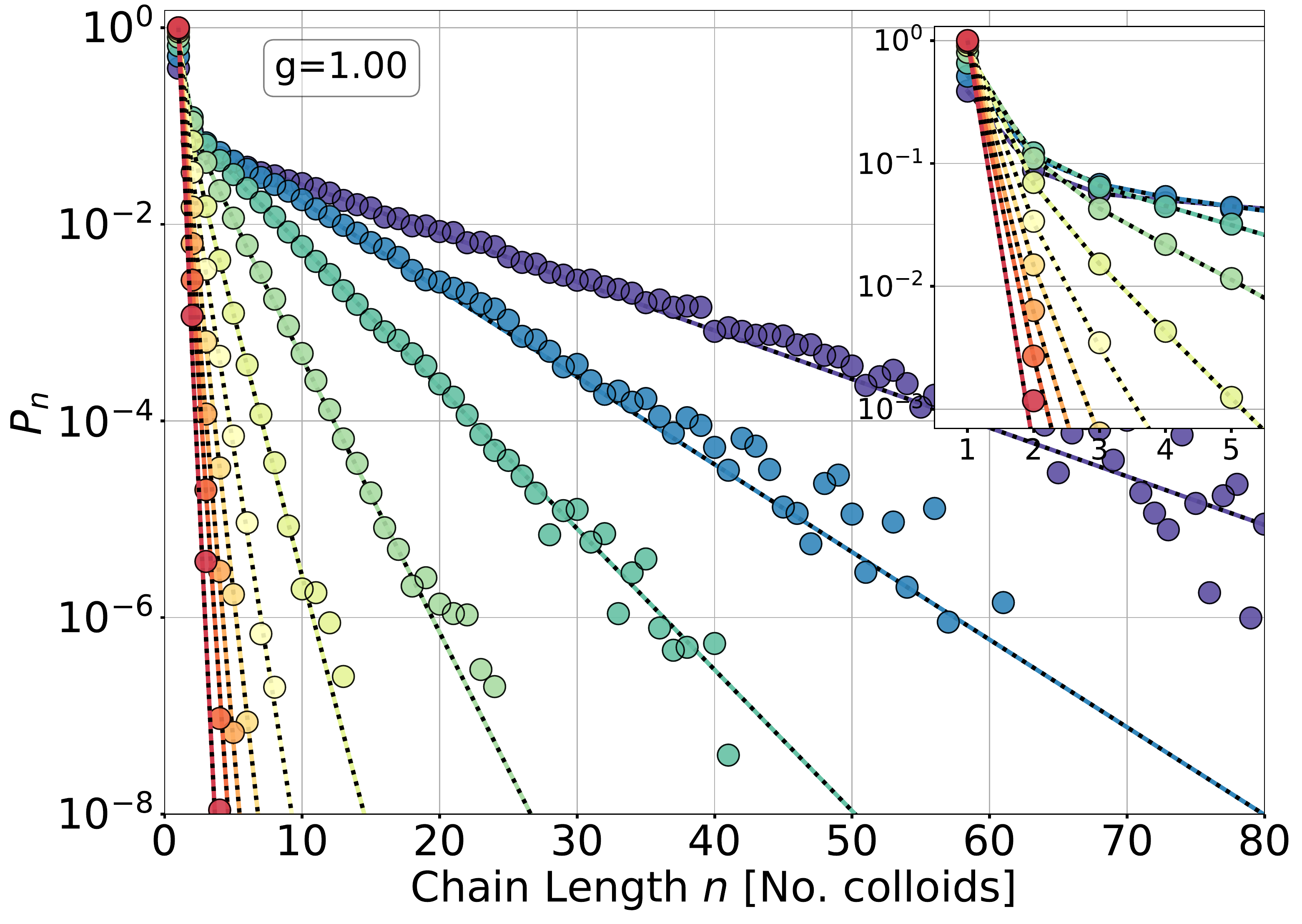}
		\includegraphics[width=0.48\textwidth]{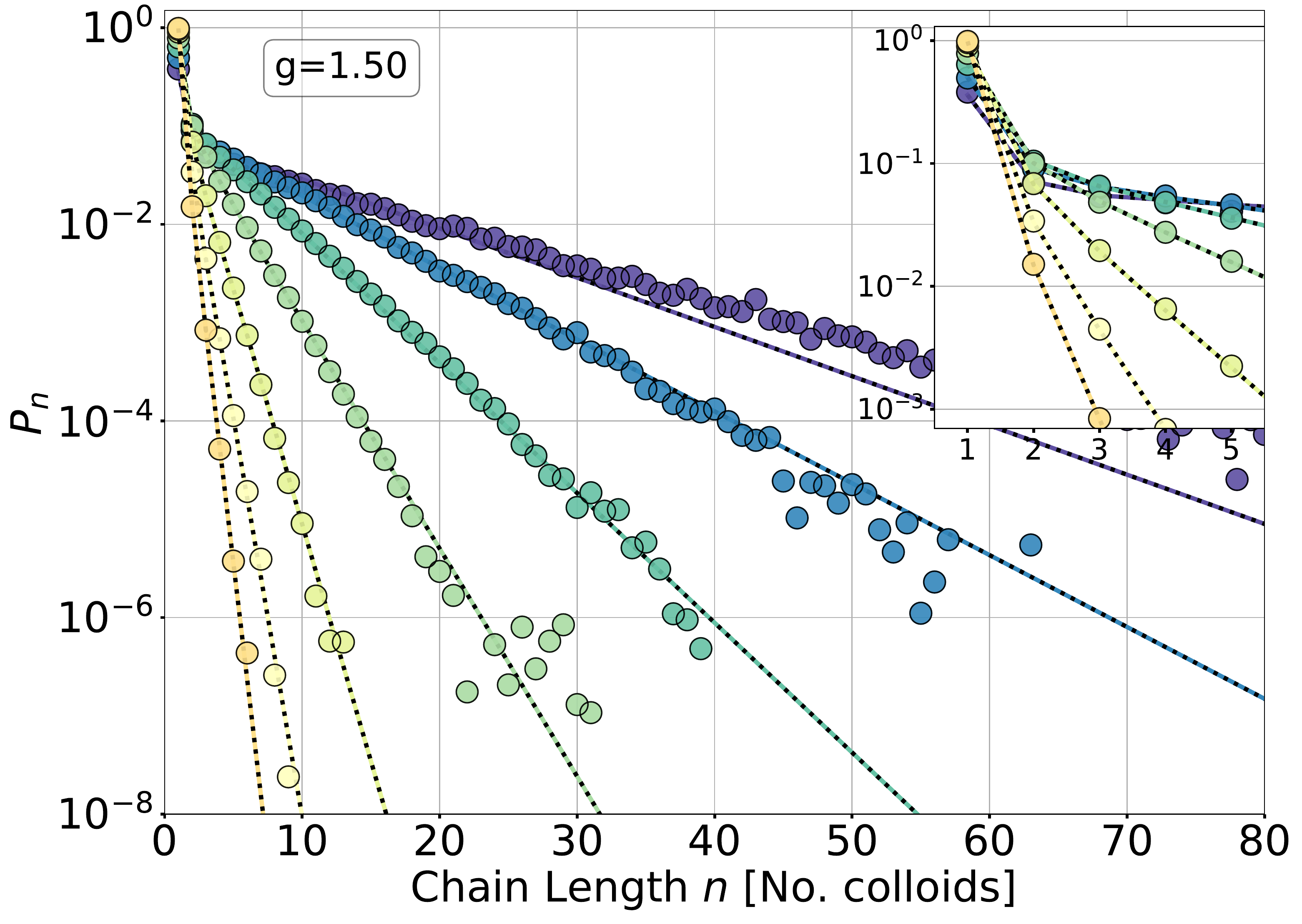}			\includegraphics[width=0.48\textwidth]{./legend_sqw.pdf}	
		\caption{ Predicted (solid lines) and simulated  (dots) chain length distributions of systems with $S_{\rm sm.}$ and $\rho$=0.255 $N/\sigma^2$. The gravitational force is varied form $g$=0.5, 1.0 and 1.5 times $F_g=-7.70 k_{\rm B}T/\sigma$. 
		} 
		\label{CLD_prediction_gravity}
	\end{figure}

	Fig.~\ref{CLD_prediction_gravity} shows simulated (dots) and predicted (solid lines) chain length distribution of the square well radial potentials with switching function  $S'=S_{\rm sm.}$, and $\theta_{\rm p}$=20$\degree$ at $\rho=0.255N/\sigma^2$ with a gravitational force of $g$=0.5, 1.0 and 1.5 times $F_g=-7.70 k_{\rm B}T/\sigma$. 
	Again,  very good agreement between predicted and simulated distributions is observed. Since the predictions included a variational $K$  up to and including chain length $l=8$ (Eq.~\ref{reaction_l-2_l-1}-\ref{nn_reactionn}), the distributions show a gradual change of the initial slope of the exponential decay, manifesting the gradual change of the reaction constant, as clearly shown in the inset.
	For $g=1.50$, in contrast, the distributions start to resemble more the quasi-2D systems and showing a sharp transition $K_2$ to $K_3$ and  $K_3\approx K_4$.  This suggest that for higher gravitation one may consider only smaller changes, and fewer  reaction constants.

\begin{figure}[t!]
	\centering
	\includegraphics[width=0.48\textwidth]{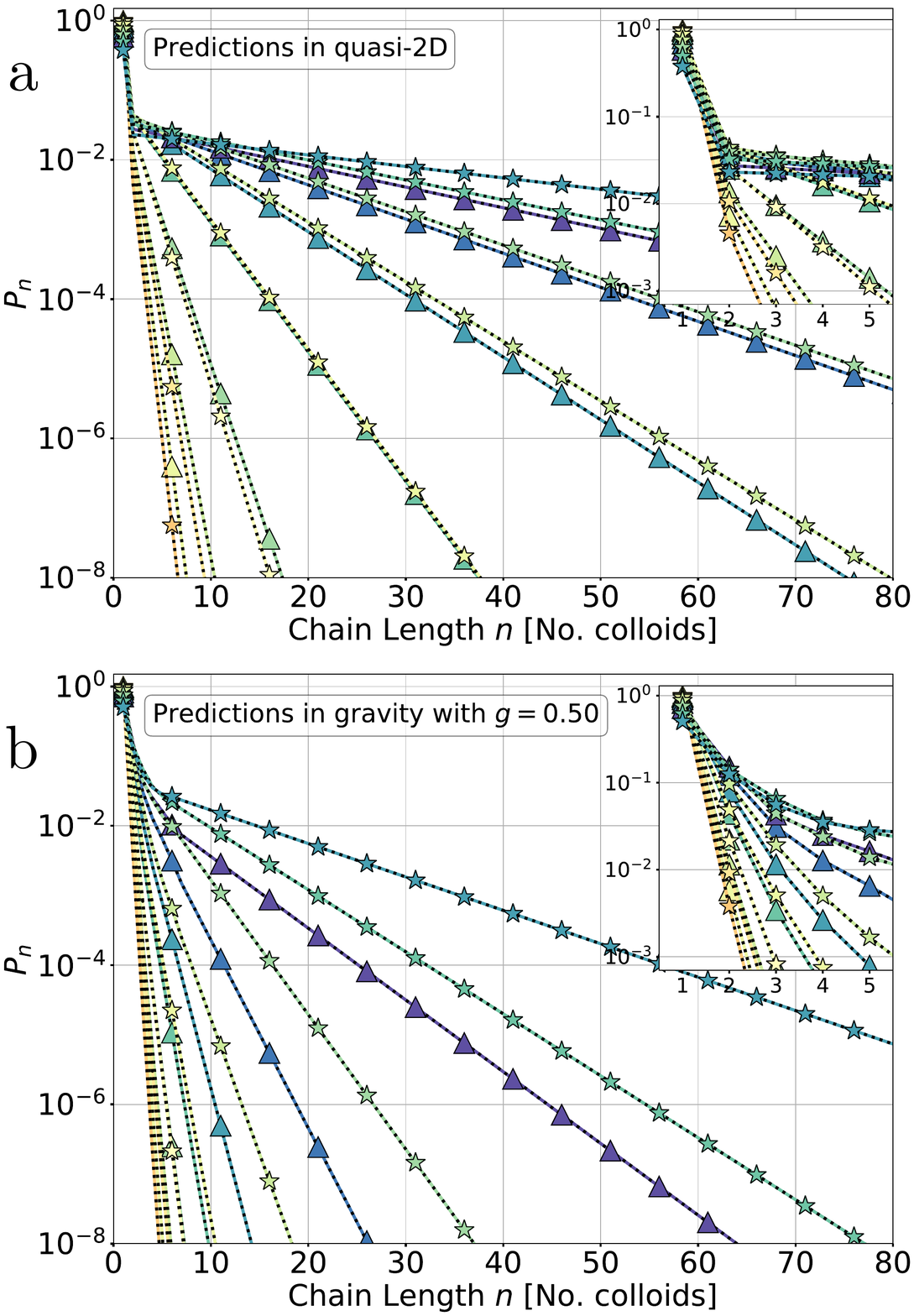}
	\caption{ The square well radial potentials with  switching functions $\theta_p=20\degree$ and $S'=S_{\rm lin.}$ (triangles), and  $\theta_p=10\degree$ and $S'=S_{\rm sm.}$ (stars) render the same distribution in quasi-2D(a).  While when exposed to a gravitational field of $g=0.50$, the distribution do not overlap(b). The color coding is the same as in Fig. \ref{rot_excess}.
	} 
	\label{compare_q2d_grav}
\end{figure}


Now that we have shown that our extended Wertheim theory is able to incorporate the effects of the gravitational field and the hard wall on the chain length distributions, we can show the role of chain's flexibility on its distribution at finite gravity.  The binding rigidity is related to excursions of the bending angle $\theta$ due to the thermal fluctuations that are on the order of $1/\beta \epsilon$. The faster $S'$ decays as function of $\theta$ (see e.g. the insets of Fig. ~\ref{sqw} and \ref{casimir}), the stiffer the bond.
When we select two systems with comparable $K_2$ and $K$, i.e. $F_{\rm rot}$ and $F_{\rm bond}$, in quasi-2D, their distributions are similar (Fig.~\ref{compare_q2d_grav}a). If these system are in a finite gravitational field instead, their statistics start to diverge significantly from each other (Fig.~\ref{compare_q2d_grav}b). We assign this divergence due to the flexibility of the chains: higher stiffness makes the alignment of the particles with the wall more prominent and reduces their reactivity. This example emphasizes the complexity of the orientation-position-dependent reactivity. As a results, there is no direct  mapping of the  statistics in quasi-2D to finite gravitational strength.
	
	\begin{figure}[t!]
		\centering
		\includegraphics[width=0.48\textwidth]{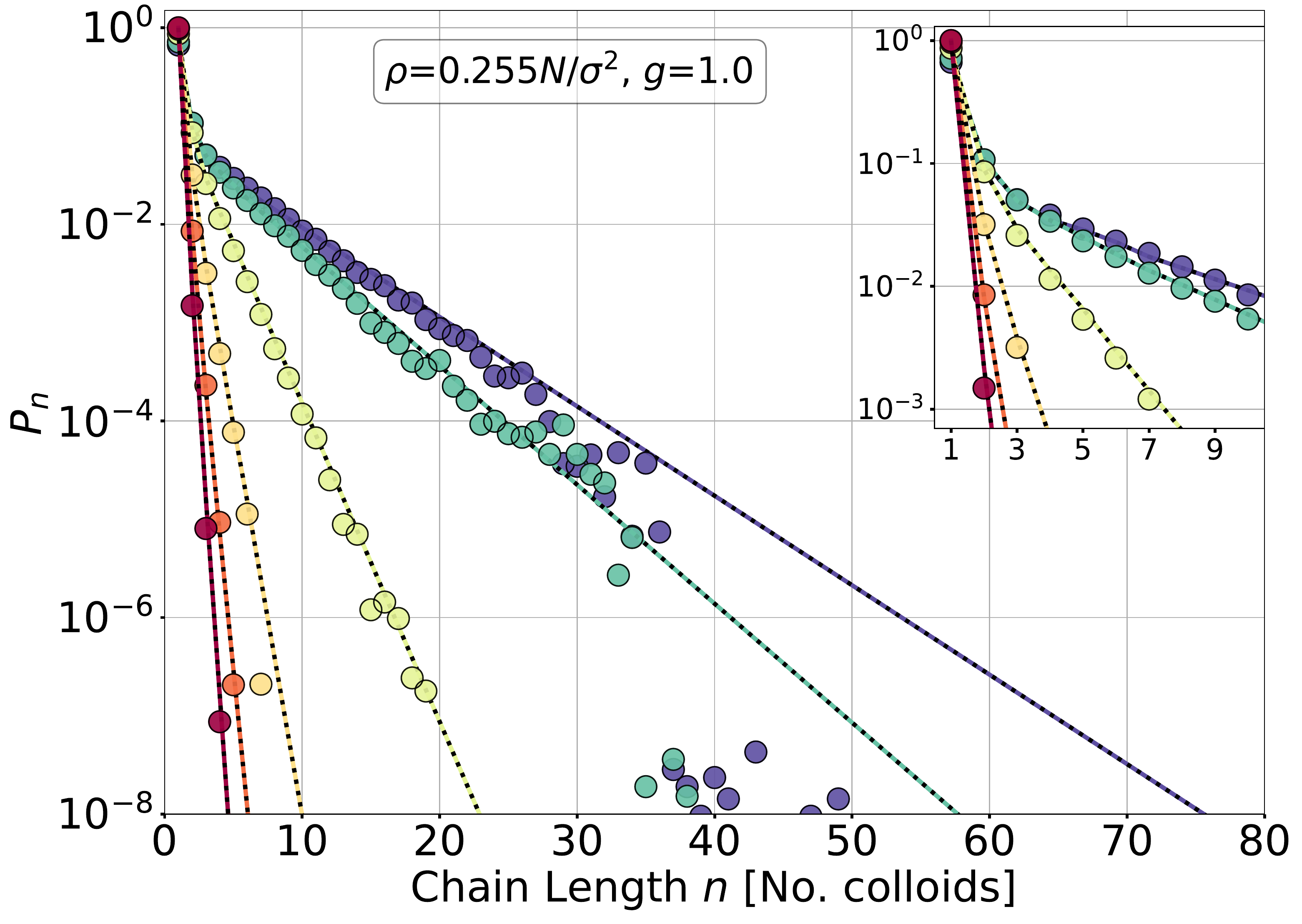}
		\includegraphics[width=0.48\textwidth]{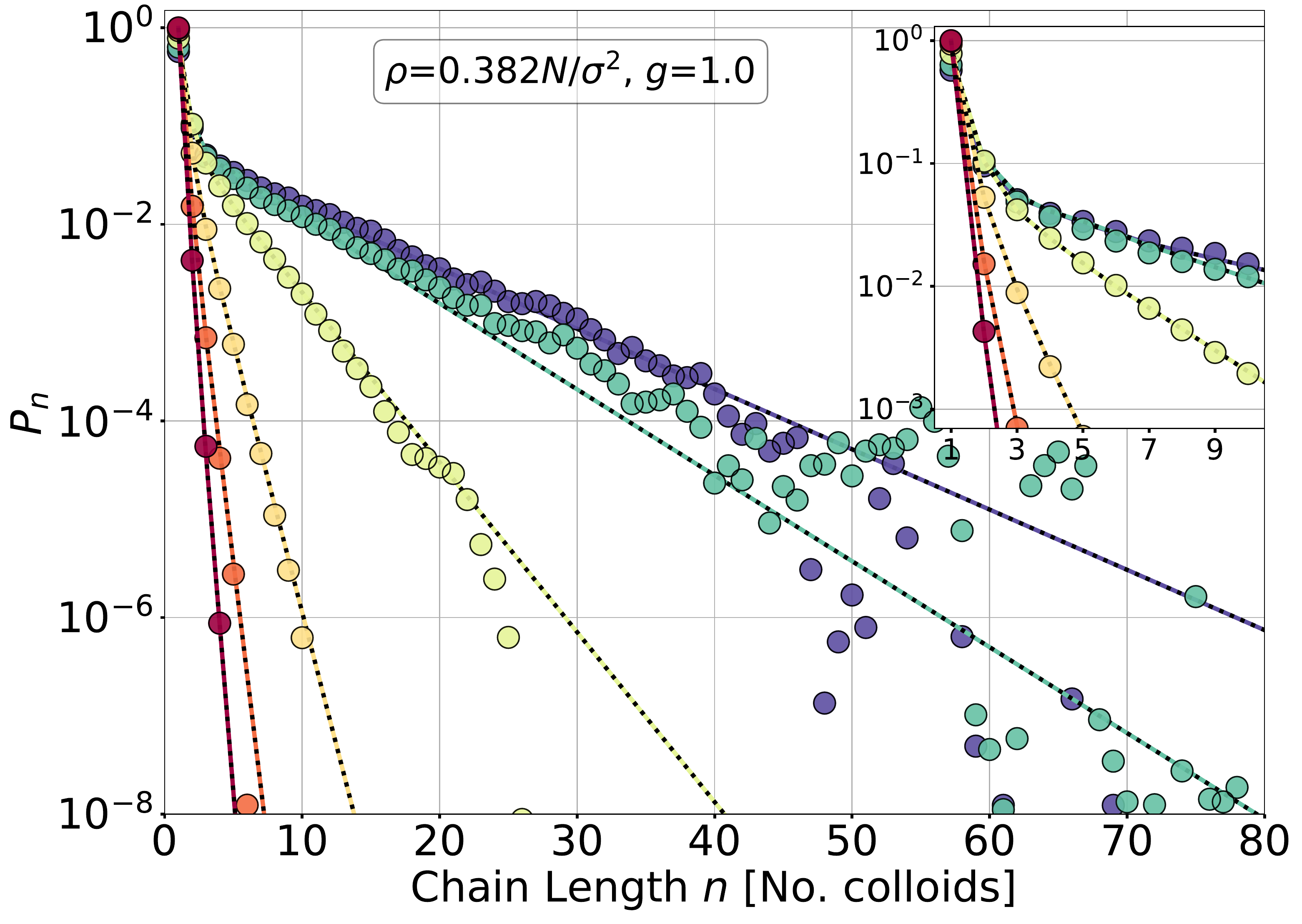}			\includegraphics[width=0.48\textwidth]{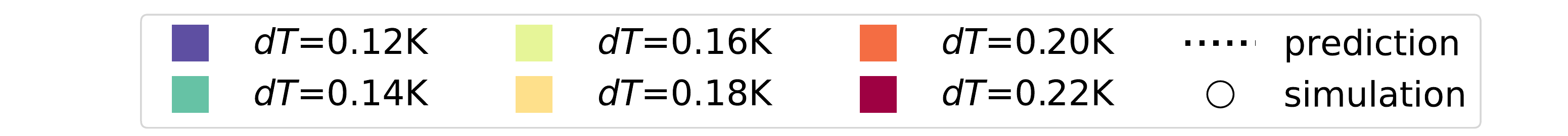}	
		\caption{ Predicted (solid lines) and simulated  (dots) chain length distributions of systems interacting via critical Casimir interactions under a realistic gravitational field at $\rho$=0.255 and 0.382 $N/\sigma^2$. 
		}
		\label{CCF_prediction_gravity}
	\end{figure}

Finally, we can apply the extended Wertheim theory to the experimentally relevant system of patchy particles interacting via the critical Casimir force shown in Fig.~\ref{CCF_prediction_gravity}. 
We can confirm that the approximations done for the RDF, namely using the WCA separation of the repulsion and attraction, and fixing the hard sphere reference diameter $d$ are sufficient. The theory predicts the distribution at various interaction strengths well, except for the strong interaction strengths at $dT=0.14$ and $0.12K$. At this region, the simulations at $dT=0.14K$ and $\rho=0.382N/\sigma^2$ show significantly longer chain lengths than predicted. 
Nematic phases are forming that promote formation of longer chains and slowing down the growth kinetics\cite{Jonas2021,Stuij2021}. This is additionally where the simulations have difficulty converging. At $dT=0.12K$ at both densities, the chains in simulation are still shorter than predicted.  Another reason for the difficulty of converging is the relatively narrow switching function in combination with the strong attraction of the critical Casimir potential. 
The current theory uses an isotropic distribution in the $(x,y)-$plane and can thus not 
 predict the enhanced reactivity due to the nematic phase.

	\section{Conclusions}

        In this work we have extended the Wertheim first-order perturbation theory to describe self-assembly of divalent particles under extreme confinement by introducing additional reaction equilibrium constants that account for the reduction of rotational and translational entropy and bond free energy of the polymerization. In the tested systems, the confinement to a monolayer of particles is created by a gravitational field that leads to sub-diameter gravitational heights and an anisotropy of particle density in the direction perpendicular to a wall. 
       
        Explicit calculation of these reaction constants from the interaction potential via the integral method allowed for a prediction of the entire chain length distribution functions that agree excellently with direct simulations of these systems. An essential part of the theory is the radial distribution function of the reference hard particle. For finite gravity, this reference hard particle distribution is computed explicitly, but only once for a certain density. The computation of the interaction parameters $\Delta$ can then be done for all densities simultaneously. 
       
        In quasi-2D, we can separate the excess rotational entropy from the bonding free energy; the results show that the patch form, size and interaction strength all play a role on the rotational free energy for non-conical potentials, while for conical potentials only the size matters. 
        Additionally, we illustrate that there is no direct or straightforward mapping of the statistics in quasi-2D onto the gravitational systems as the chain's flexibility -- thereby availability of the bonds -- defines its reactivity. This complex position-orientation dependent reactivity can be explicitly determined by our method.
        	
         As one might expect, due to the formation of nematic phases that are, in fact, also observed in experiments, our approach breaks down.  One of the assumptions, that the chains will form isotropically in the $(x,y)$-plane no longer holds. This situation is beyond the scope of the current work. 
        
        The advantage of the approach is that it only needs an approximate form of a reference radial distribution function to allow quantitative prediction of an entire range of densities and does not rely on various forms of associating density functionals. 
        Moreover, the approach allows to understand and explain the anomalous small chain concentration of self-assembly under sedimentation conditions. 
        
        Finally, we foresee that for other (molecular) self-assembling systems that are described by Wertheim’s theory in bulk, our novel extended theory can be applied to describe the system’s  behavior in extreme confinement or under an external field, e.g. in nanoslits, as the theory is validated not only for toy models but also for realistic potentials.

        \acknowledgements
        
	The authors acknowledge the funding (Grant No. 680.91.124) of the Foundation for Fundamental Research on Matter (FOM), which is part of The Netherlands Organization for Scientific Research (NWO).

	\section*{Data Availability}
	The data that supports the findings of this study are available within the article or Ref \onlinecite{Jonas2021} when indicated.

	\bibliography{MyCollection}
	%
\end{document}